\documentclass[journal]{IEEEtran}

\usepackage{amsmath,amsfonts}
\usepackage{array}
\usepackage{textcomp}
\usepackage{stfloats}
\usepackage{url}
\usepackage{verbatim}
\usepackage{graphicx}
\usepackage{cite}
\usepackage{amssymb}
\usepackage{xcolor}
\usepackage{comment}
\usepackage{multicol}
\usepackage{multirow}
\usepackage{subfigure}
\usepackage{setspace}
\usepackage{psfrag}
\usepackage{stmaryrd}
\usepackage{mathrsfs}
\usepackage{epstopdf}
\usepackage{enumerate}
\usepackage{booktabs}  
\usepackage{color}
\usepackage{dsfont,bm}
\usepackage{epsfig}
\usepackage{balance}
\usepackage{cases}% Equation label
\usepackage{enumitem}

\usepackage{algorithmic}
\usepackage{algorithm}

\usepackage{halloweenmath}
\allowdisplaybreaks[4]

\usepackage{amsthm}
\usepackage{ulem}
\newtheoremstyle{mytheostyle}{1pt}{1pt}{\itshape}{\parindent}{\bfseries}{\textnormal{\textit{:}}}{.5em}{\textit{\thmname{#1}\thmnumber{ #2}}\thmnote{ \textit{(#3)}}}
\theoremstyle{mytheostyle}

\newtheorem{remark}{Remark}
\newtheorem{theorem}{Theorem}

\newtheorem{lemma}{Lemma}

\newtheorem{corollary}{Corollary}

\newtheorem{proposition}{Proposition}

\begin{document}
\title{A Dual-Function Radar-Communication System Empowered by Beyond Diagonal Reconfigurable Intelligent Surface}
\author{Bowen Wang,~\IEEEmembership{Graduate Student Member,~IEEE,}
		Hongyu Li,~\IEEEmembership{Graduate Student Member,~IEEE,} \\
		Shanpu Shen,~\IEEEmembership{Senior Member, IEEE,}
		Ziyang Cheng,~\IEEEmembership{Member,~IEEE,}  and 
		Bruno Clerckx,~\IEEEmembership{Fellow, IEEE} 
%		\vspace{-2em}
\thanks{
Copyright (c) 2015 IEEE. Personal use of this material is permitted. However, permission to use this material for any other purposes must be obtained from the IEEE by sending a request to pubs-permissions@ieee.org.
}
\thanks{
Manuscript received 4 March 2024; revised 30 June 2024; accepted 8 August 2024.
Date of publication 22 August 2023.
The work of B. Wang and Z. Cheng was supported in part by the National Natural Science Foundation of China under Grants 62371096 and 62001084,  
in part by Sichuan Science and Technology Program under Grant 2023NSFSC1385, 
in part by the Qianyuan National Laboratory Foundation  under Grant No. KYZZ-F-02-202405-0007,
and in part by by National Key Laboratory of Science and Technology on Space Microwave under Grants No. Y23-SYSJJ-02 and HTKJ2024KL504001. 
The work of H. Li and B. Clerckx was supported by UK Research and Innovation (UKRI) grant EP/Y004086/1, EP/X040569/1, EP/Y037197/1, EP/X04047X/1, EP/Y037243/1.
The associate editor coordinating the review of this article and approving it for publication was Vahid Jamali.
(\emph{Corresponding author: Ziyang Cheng})
}
\thanks{B. Wang and Z. Cheng are with the School of Information and Communication Engineering, University of Electronic Science and Technology of China, Chengdu 611731, China. (email: bwwang@std.uestc.edu.cn, zycheng@uestc.edu.cn).}
\thanks{H. Li and B. Clerckx are with the Department of Electrical and Electronic Engineering, Imperial College London, London SW7 2AZ, U.K. (email: c.li21@imperial.ac.uk, b.clerckx@imperial.ac.uk).}
\thanks{S. Shen is with the Department of Electrical Engineering and Electronics, University of Liverpool, Liverpool L69 3GJ, U.K. (email: Shanpu.Shen@liverpool.ac.uk).}
\thanks{Color versions of one or more figures in this article are available at https://doi.org/10.1109/TCOMM.2024.3447917.}
\thanks{Digital Object Identifier 10.1109/TCOMM.2024.3447917}
}

% The paper headers
\markboth{IEEE Transactions on Communications}%
{WANG \MakeLowercase{\textit{et al.}}: A DFRC System Empowered by BD-RIS}

\maketitle

\begin{abstract}
This work focuses on the use of reconfigurable intelligent surface (RIS) in dual-function radar-communication (DFRC) systems to improve communication capacity and sensing precision, and enhance coverage for both functions. 
In contrast to most of the existing RIS aided DFRC works where the RIS is modeled as a diagonal phase shift matrix and can only reflect signals to half space, we propose a novel beyond diagonal RIS (BD-RIS) aided DFRC system. 
Specifically, the proposed BD-RIS supports the hybrid reflecting and transmitting mode, and is compatible with flexible architectures, enabling the system to realize full-space coverage and to achieve enhanced performance.
To achieve the expected benefits, we jointly optimize the transmit waveform, the BD-RIS matrices, and sensing receive filters, by maximizing the minimum signal-to-clutter-plus-noise ratio for fair target detection, subject to the constraints of the communication quality of service, different BD-RIS architectures and power budget.
To solve the non-convex and non-smooth max-min problem, a general solution based on the alternating direction method of multipliers is provided.
Numerical simulations validate the efficacy of the proposed algorithm and show the superiority of the BD-RIS aided DFRC system in terms of both communication and sensing compared to conventional RIS aided DFRC.
\end{abstract}


\vspace{-1em}
\section{Introduction}
\IEEEPARstart{I}{n recent} years, spectrum resources are becoming increasingly limited and valuable due to the exponential growth of services in wireless communications. 
Meanwhile, radar systems are competing for the same scarce sources, which motivates the emergence of the dual-function radar-communication (DFRC) technology to achieve spectrum sharing between communication and radar.
In DFRC systems, communication and radar functionalities are integrated on a common platform, which brings the benefit of enhanced spectrum efficiency while reducing power consumption and hardware costs.
Therefore, DFRC is envisioned to play an important role in emerging environment-aware applications \cite{cui2021integrating}, such as vehicular networks, environmental monitoring, and smart houses.

Due to the benefits of DFRC, plenty of technical efforts have been devoted to designing DFRC systems. 
The design methodology can be roughly divided into three categories: radar-centric design \cite{nowak2016co,hassanien2016signaling,wu2021frequency}, communication-centric design \cite{kumari2017ieee,dokhanchi2019mmwave,sturm2011waveform}, and joint waveform design \cite{zhang2021overview,liu2022integrated,cheng2023twc}. 
Radar-centric approaches utilize the radar waveform as the information carrier, where the communication symbols are embedded in conventional radar signals, such as linear frequency modulation \cite{nowak2016co} and frequency hopping \cite{wu2021frequency}. 
On the other hand, communication-centric approaches realize the radar sensing tasks by modifying existing communication protocols \cite{kumari2017ieee} and waveforms \cite{dokhanchi2019mmwave,sturm2011waveform}.
In contrast to the first two categories \cite{nowak2016co,hassanien2016signaling,wu2021frequency,kumari2017ieee,dokhanchi2019mmwave,sturm2011waveform}, the DFRC waveforms can be jointly designed to provide more design freedoms so as to enhance both functionalities \cite{zhang2021overview,liu2022integrated,cheng2023twc}.
Despite the above works \cite{nowak2016co,hassanien2016signaling,wu2021frequency,kumari2017ieee,dokhanchi2019mmwave,sturm2011waveform,zhang2021overview,liu2022integrated,cheng2023twc} achieve satisfactory sensing and communication performance, one limitation is that they rely on the line-of-sight (LoS) links between the base station (BS) and communication users/sensing targets, which, however, yields the following two issues in practice:
1) The LoS link toward sensing targets or communication users can be easily blocked by obstacles.
2) The LoS channels may suffer from severe path-loss especially for high frequencies.

To overcome these issues, a promising technology named reconfigurable intelligent surface (RIS) \cite{DiRenzo2020,SGong2019,K-KWong,QWu2019} can be leveraged.
Specifically, RIS consists of numerous passive reconfigurable scattering elements with low hardware cost and power consumption \cite{DiRenzo2020,SGong2019,K-KWong,QWu2019}. 
By properly placing and adjusting the RIS, virtual non-LoS (NLoS) links can be established to ``bypass'' obstacles, thereby compensating for the path-loss and enhance system performance.
Due to its advantages, 
RIS has been explored in various DFRC systems \cite{liu2022joint,wei2022multi,yan2022reconfigurable,sankar2022beamforming,liu2023snr,song2022cram,Hua2022Joint,wang2021joint,sankar2021joint} to enhance both the communication and sensing performance, which are classified into the following two categories.
The \textit{first} category assumes LoS links exist from BS to users and targets.
In this category, the RIS is used to compensate for the propagation loss and to improve the performance \cite{liu2022joint,wei2022multi,yan2022reconfigurable,sankar2022beamforming,liu2023snr}.
The \textit{second} category focuses on the scenario where either communication users or sensing targets are blocked by barriers. 
In this category, RIS is utilized to establish a NLoS link to bypass the barriers and thus enable DFRC \cite{song2022cram,Hua2022Joint,wang2021joint,sankar2021joint}.

The limitation of the aforementioned works \cite{liu2022joint,wei2022multi,yan2022reconfigurable,sankar2022beamforming,liu2023snr,song2022cram,Hua2022Joint,wang2021joint,sankar2021joint} is that they assume the RIS can only reflect signals towards the same side as the BS. 
In this case, both communication users and sensing targets should be located at the same side of RIS, i.e., within the same half-space, which limits the coverage and beam control flexibility of the RIS enabled DFRC system.
To address this limitation, a novel hybrid transmissive and reflective RIS, namely simultaneously transmitting and reflecting RIS (STAR-RIS) \cite{JXuHybridRIS} or intelligent omni-surface \cite{HZhang}, is proposed to support signal reflection and transmission and thus extend the coverage. 
The integration of STAR-RIS and DFRC is first studied in \cite{wang2022stars}, where the system is designed by minimizing the Cram\'er-Rao bound (CRB) for radar target estimation subject to communication constraints.
Then, a STAR-RIS is deployed at the vehicle to improve both sensing and communication performance \cite{meng2022sensing}.
Recently, the authors in \cite{zhang2023star} investigate joint uplink communication and downlink sensing, where the BS performs communication and sensing to the transmissive and reflective area in a time division duplex manner.

Although the above-mentioned STAR-RIS aided DFRC works \cite{wang2022stars,meng2022sensing,zhang2023star} have shown their advantages in enlarging coverage and boosting performance, there are several limitations of them and challenges to be solved:

\textit{First,} the achievable performance of STAR-RIS aided DFRC in \cite{wang2022stars,meng2022sensing,zhang2023star} is limited by its simple architecture that has limited wave manipulation capability.
To enhance the performance of RIS, beyond diagonal RIS (BD-RIS) \cite{SShen,li2022beyondtwc,nerini2022optimal,li2022beyond_MSM,LQC2022NDRIS} has been proposed as a generalization of conventional RIS and STAR-RIS by exploring different architectures/modes.
BD-RIS was first proposed in \cite{SShen} and has been proved to provide more controllable scattering matrices than conventional RIS.
Then, the hybrid reflective and transmissive BD-RIS is proposed in \cite{li2022beyondtwc} to achieve full-space coverage, which, in the meantime, enables single/group/fully-connected architectures based on the flexible connections among antenna ports.
It has been shown in \cite{li2022beyondtwc} that BD-RIS not only realizes full-space coverage, but achieves better performance than STAR-RIS. 
However,  previous works \cite{SShen,li2022beyondtwc,nerini2022optimal,li2022beyond_MSM,LQC2022NDRIS} have primarily concentrated on modeling and architectural design. 
Thus, the application of BD-RIS in DFRC systems and a comprehensive investigation of its advantages remain open challenges.

\textit{Second,} most existing works \cite{wang2022stars,meng2022sensing,zhang2023star} consider the simplified scenario, where the BS provides the communication service to the transmissive area, while performing sensing to the reflective area.
Besides, the previous works \cite{wang2022stars,meng2022sensing,zhang2023star} only consider sensing one target in the design with the absence of clutter sources, limiting their applicability in practical DFRC scenarios.
In real-world scenarios, both the transmissive and reflective areas often consist of multiple targets to be detected and multiple users to be served. 
Additionally, the environment typically contains clutter sources like buildings and trees, which cause strong scatterback signals that interfere with the radar sensing process. 
Therefore, it is essential to consider a more general DFRC scenario which includes the above-mentioned practical issues.

\textit{Third,} the design of STAR-RIS aided DFRC system is complicated, non-convex and hard to tackle.
Therefore, existing works \cite{wang2022stars,meng2022sensing,zhang2023star} design the transmit covariance instead of the transmit waveform for simplicity.
To align well with real-world scenarios, in this paper, we focus on the waveform design.
When considering the aforementioned general scenario, the design procedure becomes more sophisticated. 
Furthermore, the BD-RIS architectures introduce new hardware constraints, which in turn bring new challenges. 
The difficulties of waveform design in general multi-target and multi-user scenarios and the newly introduced BD-RIS constraints motivate the development of effective yet efficient algorithms.

Motivated by the above discussions, we propose a BD-RIS aided DFRC systems to achieve full-space coverage and better performance.
The main contributions of this work are summarized as follows:

\textit{First,}
we propose a general and practical BD-RIS aided DFRC system, which consists of a BD-RIS enabling the full-space coverage, multiple users, and multiple sensing targets corrupted by multiple clutter sources.
The BD-RIS divides the space into two sides and establishes virtual NLoS links for communication and sensing, where the dual-function BS (DFBS) simultaneously performs communication tasks and sensing tasks in both sides.
To avoid multi-step path-loss, we implement the radar sensing receiver on the BD-RIS for multi-target detection.
The proposed scenario is general enough to include the existing STAR-RIS aided DFRC as special cases, while it goes beyond that and includes practical factors, such as multiple targets and clutter sources.

\textit{Second,}
we formulate the unique optimization problem for the considered scenario to jointly design the transmit waveform at the DFBS, the reflective and transmissive beamforming at the BD-RIS, and matched filters at the radar sensing receiver, to maximize the minimum radar output signal-to-clutter-plus-noise ratio (SCNR), subject to the communication quality of service (QoS) requirement for downlink communications, the transmit power constraint at the DFBS, and the constraints for BD-RIS with different architectures.

\textit{Third,}
to address the complicated and non-smooth objective and the newly introduced constraints of BD-RIS, we propose to decouple the problem by the alternating direction method of multipliers (ADMM) framework, which results in a series of sub-problems.
All sub-problems are reformulated into easily handled forms and iteratively solved until convergence, in which the BD-RIS constrained sub-problems are tackled by a novel singular value decomposition (SVD) based method.

\textit{Forth,}
we provide simulation results to illustrate the performance improvement achieved by BD-RIS.
It is shown that benefiting from the high flexibility of BD-RIS, and the joint design of transmit waveform, BD-RIS, and the matched filters, the group/fully-connected BD-RISs can achieve higher radar SCNR than STAR-RISs under the same communication requirement.
It is also shown the BD-RIS can substantially improve the performance and coverage compared to the conventional RIS, which verifies the benefits of BD-RIS in manipulating the incident signal to enhance the DFRC performance.

\textit{Organization:}
Section \ref{sec:2} presents the system model of the proposed BD-RIS aided DFRC.
Section \ref{sec:3} formulates the max-min fairness problem and provides a joint design algorithm.
Section \ref{sec:4} evaluates the performance of the proposed algorithm and compares different BD-RIS architectures. 
Section \ref{sec:5} concludes this work.

\textit{Notation:}
Scalars, vectors and matrices are denoted by standard lowercase letter $a$, lower case boldface letter $\bf{a}$ and upper case boldface letter $\bf{A}$, respectively.
$\mathbb{C}^{n}$ and $\mathbb{C}^{m\times n}$ denote the $n$-dimensional complex-valued vector space and $m\times n$ complex-valued matrix space, respectively.
$(\cdot)^T$, $(\cdot)^H$, and $(\cdot)^{-1}$ denote the transpose, conjugate-transpose operations, and inversion, respectively.
$\Re \{ \cdot \}$ and $\Im \{ \cdot \}$ denote the real and imaginary part of a complex number, respectively.
$\| \cdot \|_F$ and $| \cdot |$ denote the Frobenius norm and magnitude, respectively.
$\text{Diag}(\cdot)$ denotes a diagonal matrix.
$\text{BlkDiag}(\cdot)$ denotes a block matrix such that the main-diagonal blocks are matrices and all off-diagonal blocks are zero matrices.
$\mathbf{I}_L$ indicates an $L \times L$ identity matrix.
$\jmath$ denotes imaginary unit.
$\angle(\cdot)$ represent the phase values of a matrix.
$\text{Tr}(\cdot)$ denotes the summation of diagonal elements of a matrix. 
$\left\lfloor \cdot \right\rfloor$ is the round-down operation.

\section{System Model}\label{sec:2}

\begin{figure}
	\centering
	\includegraphics[width = 0.9\linewidth]{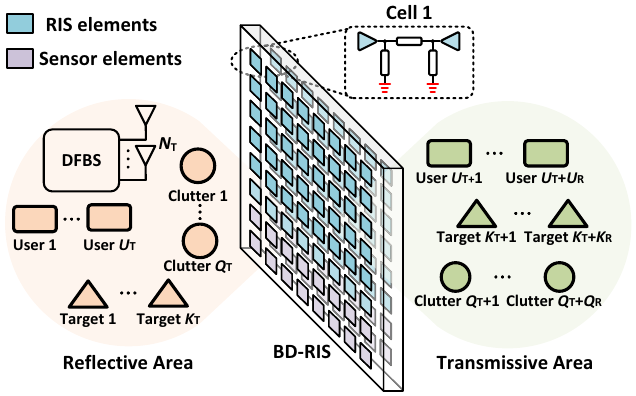}
	\caption{Illustration of a BD-RIS aided DFRC system.}
	\label{Fig_1}
\end{figure}

In this section, we present the general scenario of the proposed BD-RIS aided DFRC system, review the modeling of BD-RIS with different architectures, and establish the communication and radar models.

\subsection{BD-RIS Empowered DFRC System}
As depicted in Fig. \ref{Fig_1}, we consider a DFRC system, where an $N_\mathrm{T}$-antenna DFBS simultaneously sends communication symbols to $U$ single-antenna users and detects $K$ targets in the presence of $Q$ strong clutter sources with the assistance of an $N_\text{S}$-cell BD-RIS.
Due to the blockage or the unfavorable propagation environment, we assume the direct link between the DFBS and the users/targets does not exist.
As such, a BD-RIS working in the hybrid transmissive and reflective mode is properly placed to serve the users and illuminate targets.
To avoid multi-step path-loss, the radar sensing receiver with $N_\mathrm{R}$ antennas is installed adjacent to the BD-RIS to collect target echoes and conduct target detection tasks.
Moreover, there is a backhaul link functioning as information exchange, control and synchronization between the DFBS and the BD-RIS.

The BD-RIS divides the whole space into two half areas, i.e., the transmissive and reflective areas.
We assume that there are $U_\mathrm{T}$ single-antenna users located in the transmissive area, indexed by $u \in \mathcal{U}_\mathrm{T} = \{1 , \cdots , U_\mathrm{T}\}$, and  $U_\mathrm{R }$ single-antenna users located in the reflective area, indexed by $u \in \mathcal{U}_\mathrm{R } = \{U_\mathrm{T}+1 , \cdots , U_\mathrm{T}+U_\mathrm{R }\}$, where $\mathcal{U} = \mathcal{U}_\mathrm{T} \cup \mathcal{U}_\mathrm{R }$, $\mathcal{U}_\mathrm{T} \cap \mathcal{U}_\mathrm{R } = \emptyset$, and $U = U_\mathrm{T} + U_\mathrm{R }$.
Similarly, both reflective and transmissive areas also accommodate targets of interests, where there are $K_\mathrm{T}$/$K_\mathrm{R }$ targets located in transmissive/reflective areas with the presence of $Q_\mathrm{T}$/$Q_\mathrm{R }$ clutter sources.
Let $\mathcal{K}_\mathrm{T} = \{1 , \cdots , K_\mathrm{T}\}$, $\mathcal{K}_\mathrm{R } = \{K_\mathrm{T} + 1 , \cdots , K_\mathrm{T} + K_\mathrm{R }\}$, $\mathcal{Q}_\mathrm{T} = \{1 , \cdots , Q_\mathrm{T}\}$ and $\mathcal{Q}_\mathrm{R } = \{Q_\mathrm{T}+1 , \cdots , Q_\mathrm{T}+Q_\mathrm{R }\}$, where $\mathcal{K} = \mathcal{K}_\mathrm{T} \cup \mathcal{K}_\mathrm{R }$, $\mathcal{K}_\mathrm{T} \cap \mathcal{K}_\mathrm{R } = \emptyset$, $K = K_\mathrm{T} + K_\mathrm{R }$, $\mathcal{Q} = \mathcal{Q}_\mathrm{T} \cup \mathcal{Q}_\mathrm{R }$, $\mathcal{Q}_\mathrm{T} \cap \mathcal{Q}_\mathrm{R } = \emptyset$, and $Q = Q_\mathrm{T} + Q_\mathrm{R }$.

\subsection{BD-RIS Architecture}\label{Sec:2-B}

Based on \cite{li2022beyondtwc}, the hybrid reflective and transmissive mode is essentially based on the group-connected reconfigurable impedance network.
Each pair of antenna ports is interconnected to form a single cell, as illustrated in Fig. \ref{Fig_1}.
Within each cell, two antennas with uni-directional radiation pattern are positioned back-to-back such that each antenna covers half space. 
Following the scattering parameter network theory, the BD-RIS with hybrid reflective and transmissive mode is characterized by two matrices: transmissive matrix $\mathbf{\Phi}_\mathrm{T}\in\mathbb{C}^{N_\text{S}\times N_\text{S}}$ and reflective matrix $\mathbf{\Phi}_\mathrm{R}\in\mathbb{C}^{N_\text{S}\times N_\text{S}}$.  

Depending on the specific topology of the reconfigurable impedance network, three distinct  BD-RIS architectures have been identified \cite{li2022beyondtwc}, namely the cell-wise single/group/fully-connected (CW-SC/GC/FC) architectures, where the CW-GC architecture is the general case including CW-SC and CW-FC as two special cases.\footnote{As demonstrated in prior works \cite{SShen,li2022beyondtwc,nerini2022optimal,li2022beyond_MSM,LQC2022NDRIS}, the BD-RIS is entirely implemented using passive circuit components, illustrating that BD-RIS does not require additional power consumption.} 
To elaborate, for CW-GC BD-RIS, the entire $N_\text{S}$ cells are partitioned into $G$ groups and each group has the same size $M = N_\text{S} / G$.
The antennas within the same group are fully-connected while those of different groups are mutually independent.
Hence, the transmissive matrix $\mathbf{\Phi}_\mathrm{T}$ and reflective matrix $\mathbf{\Phi}_\mathrm{R}$ for CW-GC BD-RIS can be modeled as
\begin{equation}
	\begin{aligned}
		& \mathbf{\Phi}_{\mathrm{T}}=\text{BlkDiag}(\mathbf{\Phi}_{\mathrm{T},1},\ldots,\mathbf{\Phi}_{\mathrm{T},G}),\\
		& \mathbf{\Phi}_{\mathrm{R }}=\text{BlkDiag}(\mathbf{\Phi}_{\mathrm{R },1},\ldots,\mathbf{\Phi}_{\mathrm{R },G}),\\
		& \mathbf{\Phi}_{\mathrm{T},g}^{H}\mathbf{\Phi}_{\mathrm{T},g}+\mathbf{\Phi}_{\mathrm{R },g}^{H}\mathbf{\Phi}_{\mathrm{R },g}=\mathbf{I}_{{M}},\forall g = 1 , \cdots , G .
	\end{aligned}
	\label{eq:group_blkdiag}
\end{equation}
where ${\bm \Phi}_{\mathrm{T},g} \in {\mathbb{C}^{M \times M}}$ and ${\bm \Phi}_{\mathrm{R },g} \in {\mathbb{C}^{M \times M}}$.
Then we have the following two special cases:

\textit{Special Case 1: CW-SC BD-RIS architecture.}
In this case, we have $G=N_S$ and matrices $\mathbf{\Phi}_{\mathrm{T}}$, $\mathbf{\Phi}_{\mathrm{R }}$ are all restricted to be diagonal, i.e., $\mathbf{\Phi}_{\mathrm{T}}=\text{Diag}(\phi_{\mathrm{T},1},\ldots,\phi_{\mathrm{T},N_\text{S}})$ and $\mathbf{\Phi}_{\mathrm{R }}=\text{Diag}(\phi_{\mathrm{R },1},\ldots,\phi_{\mathrm{R },N_\text{S}})$, and satisfy
\begin{equation}
	|\phi_{\mathrm{T},i}|^{2}+|\phi_{\mathrm{R },i}|^{2}=1,\forall i = 1 , \cdots, {N_\text{S}} ,
	\label{eq:Single_connected}
\end{equation}
which conforms to the STAR-RIS constraints, indicating that the STAR-RIS is a special case of BD-RIS with CW-SC architecture \cite{HZhang,JXuHybridRIS}.

\textit{Special Case 2: CW-FC BD-RIS architecture.}
In this case, we have $G=1$ and $\mathbf{\Phi}_{\mathrm{T}}$, $\mathbf{\Phi}_{\mathrm{R }}$ are all full matrices satisfying
\begin{equation}
	\mathbf{\Phi}_{\mathrm{T}}^{H}\mathbf{\Phi}_{\mathrm{T}} + \mathbf{\Phi}_{\mathrm{R }}^{H}\mathbf{\Phi}_{\mathrm{R }} = \mathbf{I}_{N_\text{S}}.
	\label{eq:Fully_connected}
\end{equation}
Evidently, the CW-FC BD-RIS architecture offers the highest degree of freedom (DoF) among all the BD-RIS architectures by relaxing constraints at the expense of increased hardware intricacies.

\subsection{Communication Model}
In this paper, we consider a standard multiuser multiple-input single-output (MISO) downlink scenario, where the DFBS provides communication services to both transmissive and reflective areas aided by the BD-RIS. 
We assume the direct links between the DFBS and downlink users are blocked and the channel state information (CSI) is available at the DFBS.
The data symbol vector ${\bf s}_l = \left[ {\bf s}_{l}\left[ 1 \right] , \cdots , {\bf s}_{l}\left[ U \right] \right]^T \in {\mathbb C}^{U}$ contains the overall $U$ data symbols in the $l$-th time slot, which are assumed to be drawn from a standard ${\mathbb M}$ order phase-shift keying (${\mathbb M}$-PSK) modulation constellation.
Furthermore, the data symbol vector ${\bf s}_l$ is mapped to the transmit waveform $\mathbf{w}\left[l\right]\in {\mathbb C}^{N_\mathrm{T}}$ at $l$-th time slot.
Accordingly, the received signal of the $u$-th user at symbol time $t$ is 
\begin{equation}
	\begin{aligned}
		{y_u}\left( t \right) = & {e^{\jmath 2\pi {f_\mathrm{c}}t}}\sum\limits_{l \in \mathcal{L}} {{\bf{h}}_u^H{{\bf{\Phi }}_\imath}{\bf{Gw}}\left[ l \right]{\mathrm{rect}}\left( {t - l\Delta t} \right)}  \\
		& + {n_{\mathrm{c},u}}\left( t \right), \; \imath \in \{ \mathrm{T} , \mathrm{R} \} , \; \forall u \in \mathcal{U}_\imath ,
	\end{aligned}
\end{equation}
where $f_\mathrm{c}$ is the carrier frequency, $\mathcal{L} = \{1 , \cdots , L\}$ with $L$ being the number of time slots during one transmission duration, ${\bf G} \in {\mathbb C}^{N_\text{S} \times N_\mathrm{T}}$ and ${\bf h}_u \in {\mathbb C}^{N_\text{S}}$ stand for the channel matrices of the communication links DFBS$\to$BD-RIS and BD-RIS$\to$$u$-th user, $\Delta t$ stands for symbol duration, and ${n_{\mathrm{c},u}}\left( t \right)$ is the additive white Gaussian noise (AWGN).
${\rm rect}\left( t \right)$ is the rectangle window function that takes the value 1 for $t \in \left[0 , \Delta t\right]$ and 0 otherwise, which is used to illustrate discrete time slots.

By down converting the signal into baseband and sampling received signal $y_u \left( t \right)$ at the rate $f_\text{s} = 1 / \Delta t$ within the symbol duration, the discrete baseband signal at the $l$-th time slot is
\begin{equation}
	{y_u}\left[ l \right] = {\bf{h}}_u^H{{\bf{\Phi }}_\imath}{\bf{Gw}}\left[ l \right] + n_{\mathrm{c},u}\left[ l \right] , \; \imath \in \{ \mathrm{T} , \mathrm{R} \} , \; \forall u \in \mathcal{U}_\imath ,
\end{equation}
where $n_{\mathrm{c},u}\left[ l \right]$ is the AWGN with zero mean and variance $\sigma_{\mathrm{c},u}^2$.

In this work, we adopt the symbol level beamforming (SLB) technology for communication in DFRC.
Specifically, SLB technology utilizes the constructive interference (CI), which is defined as the multi-user interference (MUI) that pushes the received symbols away from the detection thresholds of the modulation constellation, to enhance the communication QoS while reducing bit error rate (BER) \cite{Li2020Tutorial,Li2018Interference}.
Here we briefly review the concept of SLB as follows. 

\begin{figure}
	\centering
	\includegraphics[width = 0.65\linewidth]{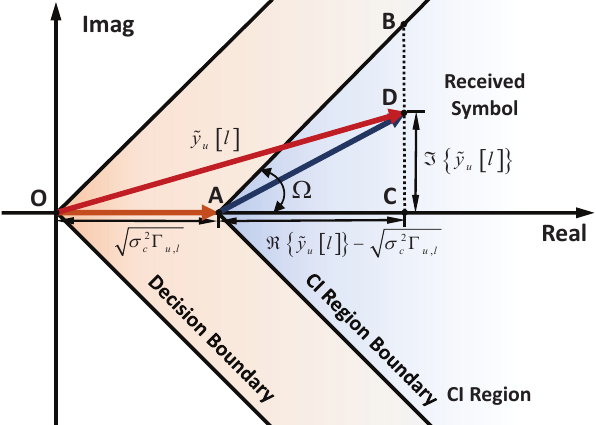}
	\caption{Description of the CI region for a QPSK symbol.}
	\label{Fig_3}
\end{figure}

Fig. \ref{Fig_3} takes quadrature-PSK (QPSK) as an example,
where point A stands for the desired symbol $\mathbf{s}_l\left[ u \right]$ with the required signal-to-noise-ratio (SNR) threshold $\Gamma_{u,l}$ of the $u$-th user, i.e., ${\overrightarrow{\text{OA}}} = \sqrt{\sigma_{\mathrm{c},u}^2 \Gamma_{u,l}} \mathbf{s}_l\left[u\right]$,
and point D is the received noise-free signal, i.e., ${\overrightarrow{\text{OD}}} = \tilde{y}_u\left[ l \right]= \mathbf{h}_u^H{\bm \Phi}_\imath\mathbf{Gw}\left[ l \right]$.
The CI region refers to a polyhedron bounded by hyperplanes parallel to decision boundaries of the constellation, which is depicted as blue-shaded area in Fig. \ref{Fig_3}.
The key of SLB is to enforce the received signal located in the CI region, which means the received signal is pushed away from decision boundaries and the SNR is guaranteed to be no less than the SNR threshold $\Gamma_{u,l}$.
To mathematically depict the SLB constraint, we project point D into the direction of $\overrightarrow{\text{OA}}$ at point C, and extend $\overrightarrow{\text{CD}}$ to intersect with the nearest boundary of CI region at point B.
Consequently, one of the criteria that specifies the location of $\overrightarrow{\text{OD}}$ in the CI region is
\begin{equation}
	\frac{|\overrightarrow{\text{CD}}|}{|\overrightarrow{\text{AC}}|} = \frac{ \left| {\Im \left\{ {{\bf{h}}_u^H{{\bf{\Phi }}_\imath}{\bf{Gw}}\left[ l \right]{e^{\jmath \angle \left( {{{\bf{s}}_u}\left[ l \right]} \right)}}} \right\}} \right| }{{\Re \left\{ {{\bf{h}}_u^H{{\bf{\Phi }}_\imath}{\bf{Gw}}\left[ l \right]{e^{\jmath \angle \left( {{{\bf{s}}_u}\left[ l \right]} \right)}}} \right\} - \sqrt {\sigma _{\mathrm{c},u}^2{\Gamma _{u,l}}} }} \le \tan  \Omega  ,
\end{equation}
where $ \Omega  = \pi / {\mathbb{M}}$ is half of the angular range of the CI resign.

\subsection{Radar Model}
For the radar function, in this paper, we focus on improving target detection performance and suppressing the clutter sources.
Specifically, to improve the sensing performance of the BD-RIS aided DFRC system, as shown in Fig. \ref{Fig_1}, we adopt a novel \textit{sensor-at-RIS} architecture \cite{shao2022target}, where the radar receiving sensors are installed adjacent to the BD-RIS to collect the echo signals.
This architecture greatly reduces the multi-step path-loss compared with the \textit{sensor-at-DFBS} architecture \cite{liu2022joint,wei2022multi,yan2022reconfigurable}.
Moreover, we consider a scenario where the radar receiver attempts to detect $K$ targets in the presence of $Q$ strong clutter sources. 
Specifically, the $k$-th target of interest is characterized by angle $\varphi_k$ and time delay $\tau_\mathrm{T}^k$, respectively.
The prior information of targets is assumed to be known to the DFBS and can be perfectly estimated in the target estimation stage \cite{li2008mimo}.\footnote{In this paper, we assume the targets are slowly moving or stay still, whose Doppler frequencies equal to zeros.}
Besides, we consider stationary clutter sources, such as buildings, and trees.
The $q$-th clutter source is characterized by angle $\vartheta_q$ and delay $\tau_\mathrm{C}^q$, respectively, and this information can be obtained from the environmental database.
Then, the backscattered signal at the radar receiver after down conversion is thus \cite{cheng2018spectrally,cui2013mimo,de2008design}\footnote{In this paper, we consider up to three-step propagations and omit additional forward and backward signals due to the negligible  received power.}
\begin{equation}
	\begin{aligned}
		\mathbf{r}_\imath(t) = & \sum\limits_{k\in \mathcal{K}_\imath}{\sum\limits_{i \in \mathcal{L}} \alpha_k \mathbf{A} (\varphi_k) {\bm{\Phi}}_\imath {\mathbf{Gw} [l] \text{rect} \left( {t - l\Delta t} - \tau_\mathrm{T}^k \right)}} \\
		& + \sum\limits_{q \in \mathcal{Q}_\imath } { \sum\limits_{l \in \mathcal{L}}{{\beta _q} {\bf{A}}\left( {{\vartheta _q}} \right){\bm{\Phi}_\imath}{\bf{Gw}}\left[ l \right]{\text{rect}}\left( {t - l\Delta t - {\tau _\mathrm{C}^q}} \right)} } \\
        & + {{\bf{n}}_\mathrm{r}}\left( t \right), \; \forall \imath \in \{ \mathrm{T} , \mathrm{R} \},
	\end{aligned}
	\label{eq:2_7}
\end{equation}
where $\alpha_k$ and $\beta_q$, respectively, denote the propagation coefficient for the $k$-th target and $q$-th clutter consisting of radar cross section (RCS) and channel propagation effects with ${\mathbb E}( {{{\left| {{\alpha _k}} \right|}^2}} ) = \zeta _k^2$ and ${\mathbb E}( {{{\left| {{\beta _q}} \right|}^2}} ) = \xi _q^2$.
${\bf{A}}\left( {{\varphi}} \right) = {{\bf{a}}_\mathrm{R }}\left( {{\varphi}} \right){\bf{a}}_\mathrm{T }^T\left( {{\varphi}} \right) \in {\mathbb C}^{N_\mathrm{R } \times N_\text{S}}$ is the effective radar channel, where ${{\bf{a}}_\mathrm{T }}\left( \varphi  \right) = \frac{1}{{\sqrt {{N_\text{S}}} }}[1, \cdots ,{e^{-j\frac{{2\pi }}{\lambda }d\left( {{N_\text{S}} - 1} \right)\sin \varphi }}]^T$ and ${{\bf{a}}_\mathrm{R }}\left( \varphi  \right) = \frac{1}{{\sqrt {{N_\mathrm{R }}} }}[1, \cdots ,{e^{-j\frac{{2\pi }}{\lambda }d\left( {{N_\mathrm{R }} - 1} \right)\sin \varphi }}]^T$ denote the the transmit and receive steering vector, respectively, with $d$ and $\lambda$ being element spacing and wavelength.
${\bf n}_\mathrm{r } \left( t \right)$ denotes AWGN.

Equation \eqref{eq:2_7} indicates that the radar receiver receives multiple path scatterback signals, while the scatterback signals before first targets echos only contain clutter sources and noise.
To maintain radar detection performance, we omit this nonconstructive scatterback signals by selecting the first target echo as the reference \cite{cheng2018spectrally,cui2013mimo,de2008design}.
Then, after sampling the received signal ${{\bf{r}}}\left( t \right)$ at $f_\text{s} = 1 / \Delta t$, the received baseband signal can be given by
\begin{equation}
	\begin{aligned}
		\mathbf{R}_\imath = & \underbrace { \sum\limits_{k \in \mathcal{K}_\imath } {\alpha _k{\mathbf{A}}( \varphi_k ){\bm{\Phi}}_\imath {\bf{GW}} {\bf J}_{r_\mathrm{T}^k} } }_{{\text{Target Echos}}} 
		+ \underbrace {\sum\limits_{q \in \mathcal{Q}_\imath} { \beta_q \mathbf{A} ( \vartheta_{q} ) {\bm\Phi}_\imath \mathbf{GW} \mathbf{J}_{{r_\mathrm{C}^q}} } }_{{\text{Clutter Returns}}}  \\
        & + {{\bf{N}}_\mathrm{r}},  \quad \forall \imath \in \{ \mathrm{T} , \mathrm{R} \},
	\end{aligned}
\end{equation}
where $\mathbf{W} = [ \mathbf{w}[1] , \cdots , \mathbf{w}[L] ] \in \mathbb{C}^{N_\mathrm{T} \times L}$ is the transmit waveform.
${\bf J}_{{r}} = [ {\bf 0}_{L \times r} , {\bf I}_{L} , {\bf 0}_{L \times \left( L_{\text{obs}} - L - r \right)} ] \in {\mathbb C}^{L \times L_{\text{obs}}} $ is the shift matrix
with $L_{\text{obs}} = L + \{ \max_k r_\mathrm{T}^k \} - \{ \min_k r_\mathrm{T}^k \}$ being the receiver observation length,
$r_\mathrm{T}^k = \lfloor {( {\tau_\mathrm{T}^k - \{ \min_{\tilde{k}} {\tau_\mathrm{T}^{\tilde{k}}} \} } ){f_\text{s}}} \rfloor $ the range ring of the $k$-th target, and $r_\mathrm{C}^q = \lfloor ( {\tau _\mathrm{C}^q - \{ \min_k{\tau_\mathrm{T}^k} \} } ) {f_\text{s}} \rfloor $ the range ring of the $q$-th clutter.
${\bf N}_\mathrm{r} = \left[ {\bf n}_\mathrm{r}\left[1\right] , \cdots , {\bf n}_\mathrm{r}\left[L\right] \right] \in {\mathbb C}^{N_\mathrm{R } \times L}$ is the Gaussian noise matrix with ${\bf n}_\mathrm{r}\left[l\right] \sim \mathcal{CN}\left(\mathbf{0}, \sigma_\mathrm{r}^2\mathbf{I}_{N_{\mathrm{R}}} \right) , \forall l$.

Finally, by performing the matched filter ${\bf U}_k \in {\mathbb C}^{N_{R} \times L_{\text{obs}}}$ to the $k$-th target at radar receiver, the $k$-th target detection problem can formulated as a binary hypothesis test \cite{cheng2018spectrally,cui2013mimo,de2008design}:
\begin{numcases}{}
	{\mathcal{H}_1^k} :  \text{Tr}\{ {\alpha _k}{\bf U}_k^H{{\bf{A}}}\left( {{\varphi _k}} \right){\bm \Phi}_{\imath}{\bf{GW}} {\bf J}_{r_\mathrm{T}^k} \}  \notag \\
    \qquad\quad + \sum\limits_{p\in\mathcal{K}_\imath,p\ne k}{ \text{Tr} \{ {\alpha _p}{\bf U}_k^H{{\bf{A}}}\left( {{\varphi _p}} \right){\bm \Phi}_{\imath}{\bf{GW}} {\bf J}_{r_\mathrm{T}^p} \}  }  \nonumber \\
	\qquad\quad + \sum\limits_{q \in \mathcal{Q}_\imath} { \text{Tr} \{ {\beta _q}{\bf U}_k^H{{\bf{A}}}\left( {{\vartheta _{q}}} \right){\bm \Phi}_{\imath}{\bf{GW}}{{\bf{J}}_{{r_\mathrm{C}^q}}} \} } + \text{Tr} \{ {\bf U}_k^H{{\bf{N}}_\mathrm{r}} \}, \nonumber \\
	{\mathcal{H}_0^k} :  \sum\limits_{p\in\mathcal{K}_\imath,p\ne k}{ \text{Tr} \{{\alpha _p}{\bf U}_k^H{{\bf{A}}}\left( {{\varphi _p}} \right){\bm \Phi}_{\imath}{\bf{GW}} {\bf J}_{r_\mathrm{T}^p}  \} } \nonumber \\
    \qquad\quad + \sum\limits_{q \in \mathcal{Q}_\imath} { \text{Tr} \{{\beta _q}{\bf U}_k^H{{\bf{A}}}\left( {{\vartheta _{q}}} \right){\bm \Phi}_{\imath}{\bf{GW}}{{\bf{J}}_{{r_\mathrm{C}^q}}}\} }  + \text{Tr}\{{\bf U}_k^H{{\bf{N}}_\mathrm{r}}\}. \nonumber
\end{numcases}
According to the above binary hypothesis test, the detection probability $P_D^k$ of the $k$-th target can be evaluated as \cite{de2008design}
\begin{equation}
	P_D^k = {\mathbb{Q}} \left( \sqrt{2{\text{SCNR}}_k} , \sqrt{-2\ln \left( P_{fa} \right) } \right),
	\label{eq:2_10}
\end{equation}
where ${\mathbb{Q}} \left( \cdot , \cdot \right)$ is the Marcum Q-function of order 1, $P_{fa}$ is the false alarm probability, and the radar output SCNR of the $k$-th target after the matched filtering is given by
\begin{equation}
	\label{eq:2_11}
	\text{SCNR}_k ( {\bf W} , {\bm \Phi}_\imath , {\bf U}_k) =  \varsigma_k^{-1}
	\mathbb{E} \{| \text{Tr}( \alpha_k \mathbf{U}_k^H \mathbf{A} ( \varphi_k ) \mathbf{\Phi}_\imath \mathbf{GW} \mathbf{J}_{r_\mathrm{T}^k} ) |^2\} ,
\end{equation}
where $\varsigma_k = \mathbb{E} \left\{ \right. \sum_{p\in\mathcal{K}_\imath, p \ne k} { {{{| {{\text{Tr}}( {{\alpha _p}{{\bf{U}}_k^H}{\bf{A}}\left( {{\varphi_p}} \right){{\bf{\Phi }}_{\imath}}{\bf{GW}}{\bf J}_{r_\mathrm{T}^p}} )} |}^2}} } + \sum_{q \in \mathcal{Q}_\imath} {{{| {{\text{Tr}}( {{\beta _q}{{\bf{U}}_k^H}{\bf{A}}\left( {{\vartheta _q}} \right){{\bf{\Phi }}_\imath}{\bf{GW}}{{\bf{J}}_{r_\mathrm{C}^q}}} )} |}^2}}  + \sigma_\mathrm{r}^2\left\| {\bf{U}}_k \right\|_F^2 \left. \right\} , \forall k \in \mathcal{K}_\imath$, $\imath \in \{ \mathrm{T} , \mathrm{R} \}$.

\section{Max-Min Fairness for BD-RIS Aided DFRC}\label{sec:3}
In this section, we first formulate the joint design problem for BD-RIS aided DFRC, followed by a general algorithm.
Finally, we propose an initialization scheme and analyze the computational complexity of the proposed algorithm.

\subsection{Problem Formulation}
Given that \eqref{eq:2_10} is strictly increasing in ${\text{SCNR}}_k$, for a specified value of false alarm probability $P_{fa}$, improving the detection probability $P_D^k$ of the $k$-th target is equivalent to maximize the radar output SCNR of the $k$-th target.
Moreover, for multiple target detection cases, beamforming design usually aims to improve the detection probability for all targets, especially for the weakest targets.
Therefore, to improve the overall target detection probability and guarantee target detection fairness, we propose to maximize the minimal radar output SCNR among the $K$ targets by jointly designing the transmit waveform ${\bf W}$, the BD-RIS matrices $\left\{{\bm \Phi}_\mathrm{T} , {\bm \Phi}_\mathrm{R}\right\}$, and radar receiver filters $\left\{ {\bf U}_k \right\}_{\forall k}$, subject to communication QoS constraints, transmit  power constraint, and BD-RIS constraints.
The joint design problem is thus formulated as\footnote{Since CW-GC architecture is a general case including both CW-SC and CW-FC cases, herein we focus on the design when the BD-RIS has CW-GC architecture.}
\begin{subequations}
	\begin{numcases}{\mathcal{P}^1}
		\mathop {\max }\limits_{{\bf{W}},{{\bf{\Phi }}_\mathrm{T}},{{\bf{\Phi }}_\mathrm{R}},\left\{{\bf{U}}_k\right\}}   \left\{ \min\limits_{\forall k} {\text{SCNR}_k} \left(  {\bf W} , {\bm \Phi}_\imath , {\bf U}_k \right) \right\} \\
		\qquad\;\; {\rm {s.t.}}  \;\; \frac{ \left| {\Im \left\{ {{\bf{\tilde h}}_u^H{\bf{w}}\left[ l \right]} \right\}} \right| }{{\Re \left\{ {{\bf{\tilde h}}_u^H{\bf{w}}\left[ l \right]} \right\} - \sqrt {\sigma _{{\rm c},u}^2{\Gamma _{u,l}}} }} \le \tan  \Omega  , \label{eq:3_12b}\\
		\qquad\qquad\; \left\| {\bf{W}} \right\|_F^2 \le E , \label{eq:3_12c}\\
		\qquad\qquad\; {\bm \Phi}_\mathrm{T} = {\rm BlkDiag}\left( {\bm \Phi}_{\mathrm{T},1} , \cdots , {\bm \Phi}_{\mathrm{T},G} \right) , \label{eq:3_12d} \\
		\qquad\qquad\; {\bm \Phi}_\mathrm{R} = {\rm BlkDiag}\left( {\bm \Phi}_{\mathrm{R},1} , \cdots , {\bm \Phi}_{\mathrm{R},G} \right) , \label{eq:3_12e} \\
		\qquad\qquad\; {\bm \Phi}_{\mathrm{T},g}^H{\bm \Phi}_{\mathrm{T},g} + {\bm \Phi}_{\mathrm{R},g}^H{\bm \Phi}_{\mathrm{R},g} = {\bf I}_{N_G}, \forall g  , \label{eq:3_12f}
	\end{numcases}
	\label{eq:3_12}%
\end{subequations}%
where ${\bf{\tilde h}}_u^H = \mathbf{h}_u^H {\bm{\Phi}}_\imath \mathbf{G} , u \in \mathcal{U}_\imath , \imath \in \{\mathrm{T} , \mathrm{R}\}$ is the equivalent channel for DFBS$\to$DB-RIS$\to u$-th user and $E$ is the transmit power.

Problem $\mathcal{P}^1$ is a challenging non-convex problem.
Particularly, the non-convexity stems from the complicated fractional SCNR expression in the objective and highly coupled optimization variables.
To simplify the joint design, in the following subsection, we propose a series of transformations and an ADMM based framework to decouple problem $\mathcal{P}^1$ into multiple more tractable sub-problems.

\subsection{Overview of Proposed Joint Design Framework}
To facilitate the joint design, we propose to re-arrange the SCNR \eqref{eq:2_11} into explicit and compact forms and give the following proposition.
\begin{proposition}\label{pro_0}
	The SCNR in \eqref{eq:2_11} shares the following three equivalent expressions
	\begin{subequations}
		\begin{align}
			{\textnormal{SCNR}}_k\left( {{\bf{W}},{{\bf{\Phi }}_\imath},{{\bf{U}}_k}} \right) & = \frac{{{{\bf{u}}_k^H}{{\bf{\Psi }}_{\mathrm{T},k}}{\bf{u}}_k}}{{{{\bf{u}}_k^H} \left({{\bf{\Psi }}_{\mathrm{C},k}} + \sigma_\mathrm{r}^2{{\bf{I}}_{{N_\mathrm{R}}L}} \right) {\bf{u}}_k}} ,\label{neq:2-13a} \\
			& = \frac{{{{\bf{w}}^H}{{\bm \Upsilon} _{\mathrm{T},k}} {\bf{w}}}}{{{{\bf{w}}^H}{{\bm \Upsilon} _{\mathrm{C},k}} {\bf{w}} + \sigma_\mathrm{r}^2\left\| {\bf{U}}_k \right\|_F^2}} , \label{neq:2-13b}\\
			& = \frac{{{{\bm \phi}}_\imath^H{{\bm{\Xi }}_{\mathrm{T},k}}{{{\bm \phi}}_\imath}}}{{{{\bm \phi}}_\imath^H{{\bm{\Xi }}_{\mathrm{C},k}}{{{\bm \phi}}_\imath} + \sigma_\mathrm{r}^2\left\| {\bf{U}}_k \right\|_F^2}} , \label{neq:2-13c}
		\end{align}
		\label{neq:2-13}%
	\end{subequations}
	where ${\bf u}_k = {\rm Vec}{\left({\bf U}_k\right)}$, ${\bf w} = {\rm Vec}{\left({\bf W} \right)}$, ${{{\bm \phi}}_\imath} = {\rm Vec}\left( {\bm \Phi}_\imath \right)$, $\{ {{\bf{\Psi }}_{\mathrm{T},k}} \}$, $\{ {{\bf{\Psi }}_{\mathrm{C},k}} \}$, $\{ {{\bm \Upsilon} _{\mathrm{T},k}} \}$, $\{ {{\bm \Upsilon} _{\mathrm{C},k}} \}$, $\{ {{\bf{\Xi }}_{\mathrm{T},k}} \}$, and $\{ {{\bf{\Xi }}_{\mathrm{C},k}} \}$ are defined in Appendix \ref{Proof_pro_1}.
\end{proposition}
\begin{IEEEproof}
	Please refer to Appendix \ref{Proof_pro_1}.
\end{IEEEproof}

Based on the above proposition, the objective in problem $\mathcal{P}^1$ is more tractable with respect to $\{\mathbf{u}_k\}_{\forall k}$, $\mathbf{w}$, or $\{\bm{\phi}_\imath\}_{\forall \imath}$. 
However, it is still difficult to find the solution to $\mathcal{P}^1$ due to non-convex and coupled constraints \eqref{eq:3_12b}, \eqref{eq:3_12c}, and \eqref{eq:3_12f}.
To tackle constraint \eqref{eq:3_12f}, we first define ${\bm \Phi}_{g} = [ {\bm \Phi}_{\mathrm{T},g}^H , {\bm \Phi}_{\mathrm{R},g}^H ]^H$
and rewrite \eqref{eq:3_12f} as ${\bm \Phi}_{g}^H{\bm \Phi}_{g}  = {\bf I}_{M}$.
Then, we introduce auxiliary variables ${\bm \Theta}_{g} = [ {\bm \Theta}_{\mathrm{T},g}^H , {\bm \Theta}_{\mathrm{R},g}^H ]^H = {\bm \Phi}_{g}$ and decouple constraint \eqref{eq:3_12f} into two separate constraints by adding the equality, which yields the following problem:\footnote{By introducing new auxiliary variables $\{{\bm \Theta}_{g}\}$, $\mathcal{P}^1$ is equivalently transformed into $\mathcal{P}^2$, effectively decoupling the constraints and facilitating the solution.}
\begin{subequations}
	\begin{numcases}{\mathcal{P}^2}
		\mathop {\max }\limits_{ \left\{{\bf{U}}_k\right\} , {\bf{W}},\left\{ {{\bf{\Phi }}_g} \right\},\left\{ {{\bf{\Theta }}_g} \right\} }   \left\{ \min\limits_{\forall k} {\text{SCNR}_k} \left(  {\bf W} , {\bm \Phi}_\imath , {\bf U}_k \right) \right\} \\
		\qquad\quad  {\text {s.t.}  \;\;\; \eqref{eq:3_12b}, \; \eqref{eq:3_12c}, \; \eqref{eq:3_12d}, \; \eqref{eq:3_12e}}  , \\
		\qquad\qquad\quad  {\bm \Theta}_{g}^H{\bm \Theta}_{g} = {\bf I}_{M} , \forall g, \label{eq:3_16c} \\
		\qquad\qquad\quad  {\bm \Phi}_{g} = {\bm \Theta}_{g} , \forall g. \label{eq:3_16d}
	\end{numcases}
	\label{eq:3_16}%
\end{subequations}
Problem $\mathcal{P}^2$ is a typical multi-variable optimization, which could be solved based on the ADMM framework using block coordinate descent (BCD) methods. 
To facilitate ADMM, we place the equality constraints ${\bm \Phi}_{g} = {\bm \Theta}_{g}, \forall g$ into the objective function, and obtain the augmented Lagrangian (AL) as
\begin{equation}
	\begin{aligned}
		&  {\mathcal{L}} \left( \left\{{\bf{U}}_k  \right\}  , {\bf{W}},\left\{{{\bf{\Phi }}_{g}}\right\}, \left\{ {\bm \Theta}_{g} \right\}\right)\\
		&  = - \{ \min\limits_{\forall k} {\text{SCNR}_k} \left(  {\bf W} , {\bm \Phi}_\imath , {\bf U}_k \right) \} + \sum\limits_{g=1}^{G}{ \mathbb{I}_g({\bm \Phi}_{g} , {\bm \Theta}_{g} , {\bm \Lambda}_g ) } ,
	\end{aligned}
	\label{eq:3_17}
\end{equation}
where $\mathbb{I}_g({\bm \Phi}_{g} , {\bm \Theta}_{g} , {\bm \Lambda}_g ) = {\Re \left\{ {\rm Tr} \left( {\bm \Lambda}_g^H \left( {\bm \Phi}_{g} - {\bm \Theta}_{g} \right) \right) \right\}} + \frac{\varrho}{2} \left\| {{\bm \Phi}_{g} - {\bm \Theta}_{g} } \right\|_F^2$,
$ {\bm \Lambda}_g \in {\mathbb C}^{2 M \times M }, \forall g$ are dual variables associated with ${\bm \Phi}_{g}\!=\!{\bm \Theta}_{g}$,
and $ \varrho \ge 0$ is the corresponding penalty parameter.
Replacing the original objective function with AL function \eqref{eq:3_17}, we obtain the AL minimization problem as
\begin{subequations}
	\begin{numcases}{\mathcal{P}_{\rm AL}^2}
		\mathop {\min }\limits_{ \left\{{\bf{U}}_k\right\} , {\bf{W}},\left\{ {{\bf{\Phi }}_g} \right\},\left\{ {{\bf{\Theta }}_g} \right\} }   {\mathcal{L}}\left( \left\{{\bf{U}}_k  \right\} , {{\bf{W}},\left\{ {{\bf{\Phi }}_{g}} \right\}, }  \left\{ {\bm \Theta}_{g} \right\}\right)  \\
		\qquad\quad\;  {\rm {s.t.}}  \quad\quad\quad \eqref{eq:3_12b}-\eqref{eq:3_12e}, \eqref{eq:3_16c} .
	\end{numcases}
	\label{eq:3_18}%
\end{subequations}
Now, the ADMM framework is constructed as follows, where the superscript of notations refers to the iteration index:  
\begin{subequations}
	\begin{align}
		{\bf U}_k^{ n+1 }  & = \arg \min\limits_{{\bf U}_k}{\mathcal{L}}\left( \left\{{\bf{U}}_k  \right\}, {{\bf{W}}^n, \left\{ {{\bf{\Phi }}_{g}^{n}} \right\} , } \left\{ {\bm \Theta}_{g}^n \right\}\right) \label{eq:3_19a} \\
		{\bf W}^{ n+1 } & = \arg \min\limits_{{\bf W}} {\mathcal{L}}\left( \left\{{\bf{U}}_k^{n+1}  \right\}, {{\bf{W}},\left\{ {{\bf{\Phi }}_{g}^{n}} \right\}, }  \left\{ {\bm \Theta}_{g}^n \right\}\right) \notag \\
		& \qquad\quad\; {\text{s.t.}} \; \eqref{eq:3_12b}, \eqref{eq:3_12c}. \\
		\left\{ {\bm \Phi}_{g}^{ n+1 } \right\} & = \arg \min\limits_{{\bm \Phi}_{g}}  {\mathcal{L}}\left( \left\{{\bf{U}}_k^{n+1}  \right\} , {{\bf{W}}^{n+1},\left\{ {\bf{\Phi }}_g \right\}, }  \left\{ {\bm \Theta}_{g}^n \right\}\right) \notag \\
		& \qquad\quad\; {\text{s.t.}} \; \eqref{eq:3_12b}, \eqref{eq:3_12d}, \eqref{eq:3_12e}. \\
		\left\{ {\bm \Theta}_{g}^{ n+1 } \right\} &  = \arg \min\limits_{{\bm \Theta}_{g}} {\mathcal{L}}\left( \left\{{\bf{U}}_k^{n+1}  \right\} , {{\bf{W}}^{n+1},\left\{ {{\bf{\Phi }}_{g}^{n+1}} \right\}, }  \left\{ {\bm \Theta}_{g} \right\}\right) \notag \\
		& \qquad\quad\; {\text{s.t.}} \; \eqref{eq:3_16c},  \label{eq:3_19d}\\
		{\bm \Lambda}_g^{n+1}  &  = {\bm \Lambda}_g^{n} +{\varrho} \left( {\bm \Phi}_{g}^{n+1} - {\bm \Theta}_{g}^{n+1} \right). \label{eq:3_19e}
	\end{align}
\end{subequations}
Variables \eqref{eq:3_19a} to \eqref{eq:3_19e} are successively updated by solving corresponding sub-problems until some stopping conditions are reached.
In the following subsection\footnote{When introducing solutions to sub-problems, we omit the superscript of notations for conciseness unless otherwise stated.}, we will elaborate on the solutions to sub-problems \eqref{eq:3_19a} to \eqref{eq:3_19d}.

\subsection{Solutions to Sub-problems}
\subsubsection{Sub-problem w.r.t ${\bf U}_k$}
Given other variables, the optimization problem for updating ${\bf U}_k$ can be expressed as
\begin{equation}
    {\mathcal{P}^2_{\text{AL},\left\{{\bf U}_k\right\} }}
    \left\{ 
        \mathop { \max } \limits_{ \left\{{\bf{U}}_k\right\} }   \Big\{ \min\limits_{\forall k} \frac{ {{{\bf{u}}_k^H}{{\bf{\Psi }}_{{\rm T},k}}{\bf{u}}_k} }{ {{{\bf{u}}_k^H}\left({{\bf{\Psi }}_{{\rm C},k}} + \sigma _\mathrm{r}^2{{\bf{I}}_{{N_\mathrm{R}}L}} \right){\bf{u}}_k} } \Big\} 
    \right. .
\end{equation}
It can be observed that ${\mathcal{P}^2_{\text{AL},\left\{{\bf U}_k\right\}}}$ is an unconstrained optimization problem and has $K$ separable objective functions, each of which has the following form
\begin{equation}
	    \max\limits_{{\bf U}_k} \quad \frac{{{{\bf{u}}_k^H}{{\bf{\Psi }}_{{\rm T},k}}{\bf{u}}_k}}{{{{\bf{u}}_k^H}\left({{\bf{\Psi }}_{{\rm C},k}} + \sigma _\mathrm{r}^2{{\bf{I}}_{{N_\mathrm{R}}L}} \right){\bf{u}}_k}} , \forall k .
	\label{eq:3_21}%
\end{equation}
Problem \eqref{eq:3_21} is a classical generalized fractional quadratic optimization problem, whose optimal solution can be obtained  by taking the generalized eigenvalue decomposition as \cite{cheng2018spectrally}
\begin{equation}
	 {\bf u}_k = {\text{EIG}} \left( \left\{ {{\bf{\Psi }}_{{\rm C},k}} + \sigma_\mathrm{r}^2{{\bf{I}}_{{N_\mathrm{R}}L}} \right\}^{-1} \times {{\bf{\Psi }}_{{\rm T},k}} \right) , \forall k.
	 \label{eq:3_22}
\end{equation}
where ${\rm EIG}\left( \cdot \right)$ denotes the principal eigenvector, which returns the eigenvector corresponding to the maximum eigenvalue.

\subsubsection{Sub-problem w.r.t ${\bf W}$}
Given other variables, the optimization problem for updating ${\bf W}$ can be expressed as
\begin{subequations}
	\begin{numcases}{\mathcal{P}^2_{{\rm AL},{\bf W}}}
		\mathop {\max }\limits_{{\bf{W}}}   \left\{ \min\limits_{\forall k} \frac{{{{\bf{w}}^H}{{\bm \Upsilon} _{{\rm T},k}} {\bf{w}}}}{{{{\bf{w}}^H}{{\bm \Upsilon} _{{\rm C},k}} {\bf{w}} + \sigma_\mathrm{r}^2\left\| {\bf{U}}_k \right\|_F^2}} \right\} \\
		\;\; {\rm {s.t.}}  \;\; \frac{ \left| {\Im \left\{ {{\bf{\tilde h}}_u^H{\bf{w}}\left[ l \right]} \right\}} \right| }{{\Re \left\{ {{\bf{\tilde h}}_u^H{\bf{w}}\left[ l \right]} \right\} - \sqrt {\sigma _{\mathrm{c},u}^2{\Gamma _{u,l}}} }} \le \tan  \Omega  , \label{eq:3_23b} \\
		\qquad\; \left\| {\bf{W}} \right\|_F^2 \le E . \label{eq:3_23c}
	\end{numcases}
	\label{eq:3_23}%
\end{subequations}
The following theorem provides the solution to problem $\mathcal{P}_{\mathrm{AL},\mathbf{W}}^2$.
\begin{theorem}\label{nthe:1}
	Problem $\mathcal{P}_{\mathrm{AL},\mathbf{W}}^2$ can be transformed into a convex second-order cone programming (SOCP) problem, which can be globally solved by the interior point method (IPM).
\end{theorem}
\begin{IEEEproof}
	Please refer to Appendix \ref{Proof_nthe_1}.
\end{IEEEproof}

\subsubsection{Sub-problem w.r.t $\left\{ {\bm \Phi}_{g} \right\}$}
Given other variables, the sub-problem for updating $\left\{ {\bm \Phi}_{g} \right\}$ is
\begin{subequations}
	\begin{numcases}{\mathcal{P}^2_{{\text{AL}}, {\left\{ {\bm \Phi}_{g} \right\}}}}
		\mathop {\min }\limits_{{{\bf{\Phi }}_{g}}}  - \left\{ \min\limits_{\forall k} \frac{{{{\bm \phi}} _\imath^H{{\bm{\Xi }}_{{\rm T},k}}{{{\bm \phi}}_\imath}}}{{{{\bm \phi}}_\imath^H{{\bm{\Xi }}_{{\rm C},k}}{{{\bm \phi}}_\imath} + \sigma_\mathrm{r}^2\left\| {\bf{U}}_k \right\|_F^2}}  \right\} \notag \\
        \qquad\;\; + \sum\limits_{g=1}^{G}{  \mathbb{I}_g({\bm \Phi}_{g} , {\bm \Theta}_{g} , {\bm \Lambda}_g ) } \label{eq:3_31a} \\
		{\rm {s.t.}}  \;\; \frac{{\left| {\Im \left\{ {{\text{Tr}}\left\{ {{{{\bf{\bar H}}}_{u,l}}{{\bf{\Phi }}_\imath}} \right\}} \right\}} \right|}}{{\Re \left\{ {{\text{Tr}}\left\{ {{{{\bf{\bar H}}}_{u,l}}{{\bf{\Phi }}_\imath}} \right\}} \right\} - \sqrt {\sigma _{{\rm c},u}^2{\Gamma _{u,l}}} }} \nonumber \\
		\qquad \qquad \quad \le \tan \Omega , \forall u \in \mathcal{U}_\imath , \imath \in \{{\rm T} , {\rm R}\}, \label{eq:3_31b} \\
		\quad\;\;\; {\bm \Phi}_\mathrm{T} = {\rm BlkDiag}\left( {\bm \Phi}_{\mathrm{T},1} , \cdots , {\bm \Phi}_{\mathrm{T},G} \right) ,  \label{eq:3_31c}\\
		\quad\;\;\; {\bm \Phi}_\mathrm{R} = {\rm BlkDiag}\left( {\bm \Phi}_{\mathrm{R},1} , \cdots , {\bm \Phi}_{\mathrm{R},G} \right) ,  \label{eq:3_31d}
	\end{numcases}
	\label{eq:3_31}%
\end{subequations}
where ${{{\bf{\bar H}}}_{u,l}} = {e^{\jmath \angle \left( {{{\bf{s}}_u}\left[ l \right]} \right)}}{\bf{Gw}}\left[ l \right]{\bf{h}}_u^H$.
The following theorem provides the solution to problem $\mathcal{P}_{\mathrm{AL}, {\left\{ {\bm \Phi}_{g} \right\}}}^2$.
\begin{theorem}\label{nthe:2}
	Problem $\mathcal{P}_{\mathrm{AL}, {\left\{ {\bm \Phi}_{g} \right\}}}^2$ can be transformed into a convex SOCP, which can be solved by IPM.
\end{theorem}
\begin{IEEEproof}
	Please refer to Appendix \ref{Proof_nthe_2}.
\end{IEEEproof}

\subsubsection{Sub-problem w.r.t $\left\{ {\bm \Theta}_{g} \right\}$}
Given the other variables, the sub-problem for updating $\left\{ {\bm \Theta}_{g} \right\}$ is
\begin{subequations}
	\begin{numcases}{\mathcal{P}^2_{{\text{AL}}, {\left\{ {\bm \Theta}_{g} \right\}}}}
		\mathop {\min }\limits_{{{\bf{\Phi }}_{g}}}  \sum\limits_{g=1}^{G}{ \mathbb{I}_g({\bm \Phi}_{g} , {\bm \Theta}_{g} , {\bm \Lambda}_g ) }  \\
		{\rm {s.t.}}  \;\; {\bm \Theta}_{g}^H{\bm \Theta}_{g} = {\bf I}_{M} , \forall g,
	\end{numcases}
	\label{eq:3_39}%
\end{subequations}
Problem $\mathcal{P}^2_{{\text{AL}}, {\left\{ {\bm \Theta}_{g} \right\}}}$ can be split into $G$ sub-problems, each of which has the following form
\begin{subequations}
	\begin{numcases}{\mathcal{P}^{2-1}_{{\text{AL}}, {{\bm \Theta}_{g}}}}
		\mathop {\min }\limits_{{{\bf{\Phi }}_{g}}}  \mathbb{I}_g({\bm \Phi}_{g} , {\bm \Theta}_{g} , {\bm \Lambda}_g ) \label{eq:3_40a} \\
		{\rm {s.t.}}  \;\; {\bm \Theta}_{g}^H{\bm \Theta}_{g} = {\bf I}_{M}. \label{eq:3_40b}
	\end{numcases}
	%	\label{eq:3_31}%
\end{subequations}
Now, the remaining challenge of solving problem $\mathcal{P}_{\mathrm{AL}, {{\bm \Theta}_{g}}}^{2-1}$ lies in the orthogonal constraint \eqref{eq:3_40b}.
A closed-form solution of problem $\mathcal{P}_{\mathrm{AL}, {{\bm \Theta}_{g}}}^{2-1}$ can be obtained based on the following theorem.
\begin{theorem}\label{the:1}
	With the orthogonal constraint \eqref{eq:3_40b}, the optimal solution for ${\bm \Theta}_{ g}$ is given by
	\begin{equation}
		{\bm \Theta}_{ g} = {\bf{B}}_g\left[ {{{\bf{I}}_{M \times M}},{{\bf{0}}_{M \times M}}} \right]{{\bf{D}}_g^H} ,
	\end{equation}
	where ${{\bf{B}}_g}{{\bf{\Sigma }}_g}{\bf{D}}_g^H = {{\bf{\Lambda }}_g} + {\varrho }{{\bf{\Phi }}_{g}}$ is the singular value decomposition (SVD)  of ${{\bf{\Lambda }}_g} + {\varrho }{{\bf{\Phi }}_{g}}$.
\end{theorem}
\begin{IEEEproof}
	Please refer to Appendix \ref{Proof_the_1}.
\end{IEEEproof}

Based on the above derivations, the procedure of the above ADMM based algorithm is summarized in Algorithm \ref{alg:1}.

\begin{algorithm}[t]
	\caption{Max-Min Fairness for BD-RIS Aided DFRC.}
	\label{alg:1}
	\begin{algorithmic}[1]
		\REQUIRE $\mathbf{h}_{u},\forall u$, $\mathbf{G}$, $\varrho$ and system parameters.
		%		\STATE {Calculate $\mathbf{X}_\mathrm{t/r}$, $\mathbf{Y}$, $\mathbf{Z}_\mathrm{t/r}$ by (\ref{eq:XYZ})}.
		\STATE{Initialize ${\left\{ {\bf U}_k^0 \right\}}$, ${\bf W}^0$, ${\bm \Phi}_\mathrm{T}^0$, and ${\bm \Phi}_\mathrm{T}^0$.}
		\STATE {Set $n = 1$.}
		\REPEAT
		\STATE {Calculate radar receive filters $\left\{{\bf U}_k^n\right\}$ by \eqref{eq:3_22} in parallel.}
		\STATE {Update transmit waveform ${\bf W}^n$ by \textbf{\textit{Theorem \ref{nthe:1}}}.}
		\STATE {Compute BD-RIS  matrices ${\bm \Phi}_{\mathrm{T}}^n$ and ${\bm \Phi}_{\mathrm{R}}^n$ by \textbf{\textit{Theorem \ref{nthe:2}}}.}
		\STATE {Obtain auxiliary variables $\left\{{\bm \Theta}_{g}^n\right\}$ by \textbf{\textit{Theorem \ref{the:1}}}. }
		\STATE {Update dual variables $\left\{ {\bm \Lambda}_{g}^n \right\}$ by \eqref{eq:3_19e}.}
		\STATE {$n = n + 1$.}
		\UNTIL {convergence.}
		\STATE {Return $\left\{ {\bf U}_k^n \right\}$, ${\bf W}^n$, ${\bm \Phi}_\mathrm{T}^n $ and ${\bm \Phi}_\mathrm{R}^n$.}
		\ENSURE $\left\{ {\bf U}_k^\star \right\} = \left\{ {\bf U}_k^n \right\}$, ${\bf W}^\star = {\bf W}^n$, ${\bm \Phi}_\mathrm{T}^\star = {\bm \Phi}_\mathrm{T} ^n$, ${\bm \Phi}_\mathrm{R}^\star = {\bm \Phi}_\mathrm{R} ^n$.
	\end{algorithmic}
\end{algorithm}

\subsection{Initialization Scheme}

Given that the ADMM procedure is usually sensitive to initial values, we present a 2-step initialization strategy to accelerate the convergence. 

\textit{Step 1:} Since it is not that straightforward to quickly find proper $\mathbf{\Phi}_{\mathrm{T}}$ and $\mathbf{\Phi}_{\mathrm{R}}$, we randomly generate $\mathbf{\Phi}_{\mathrm{T}}$ and $\mathbf{\Phi}_{\mathrm{R}}$, which satisfy the BD-RIS constraints.

\textit{Step 2:} With initialized ${\bm \Phi}_\mathrm{T}$ and ${\bm \Phi}_\mathrm{R}$, we obtain the cascaded channel ${\bf{\tilde h}}_u^H \left( {\bm \Phi}_\imath \right) = {\bf{h}}_u^H{{\bf{\Phi }}_\imath}{\bf{G}}, \imath \in \{ \mathrm{T} , \mathrm{R} \}$ for the communication link. 
To provide a feasible and ``good'' initial point satisfying the constraint \eqref{eq:3_12b}, we initialize the transmit waveform ${\bf W}$ by solving the following QoS-constrained problem
	\begin{equation}
		\begin{aligned}
			\max\limits_{{\bf W} , \Gamma} & \quad \Gamma \\
			{\text{s.t.}} \; & \quad \Re \left\{ {{\bf{\bar h}}_{u,1}^H\left( {{{\bf{\Phi }}_\imath}} \right){\bf{w}}\left[ l \right]} \right\} \ge \\
            & \qquad \qquad \sqrt {\sigma _{\mathrm{c},u}^2{\Gamma}} \sin \Omega ,\forall u \in \mathcal{U}_\imath , \imath \in \left\{ \mathrm{T,R} \right\}, l, \\
			& \quad \Re \left\{ {{\bf{\bar h}}_{u,2}^H\left( {{{\bf{\Phi }}_\imath}} \right){\bf{w}}\left[ l \right]} \right\} \ge \\
            & \qquad \qquad \sqrt {\sigma _{\mathrm{c},u}^2{\Gamma}} \sin \Omega ,\forall u \in \mathcal{U}_\imath , \imath \in \left\{ \mathrm{T,R} \right\}, l, \\
			& \quad \left\| {\bf W} \right\|_F^2 \le E,
		\end{aligned}
		\label{eq:3_42}
	\end{equation}
which is a convex problem and can be efficiently solved by many numerical approaches \cite{boyd2004convex}.

\begin{remark}
	It should be noted that the update of the radar receiver filters $\{\mathbf{U}_k\}$ is not included in the initialization stage for the following reasons:
	\textit{1):} The purpose of the initialization scheme is to find an initial point that satisfies the constraints \eqref{eq:3_12b}-\eqref{eq:3_12f}, which are irrelevant to $\{\mathbf{U}_k\}$.
	\textit{2):} The update of $\{\mathbf{U}_k\}$ is the first step in Algorithm \ref{alg:1}. Once the transmit waveform $\mathbf{W}$ and BD-RIS matrices $\left( {\bm \Phi}_\mathrm{T} , {\bm \Phi}_\mathrm{R} \right)$ are properly initialized, the $\{\mathbf{U}_k\}$ can be optimally updated by \eqref{eq:3_22}, without the value of $\{\mathbf{U}_k\}$ in previous rounds.
\end{remark}

\subsection{Complexity Analysis}

We provide a broad complexity analysis for Algorithms \ref{alg:1}, which is summarized as follows
\subsubsection{Initialization} The main computational complexity of this stage comes from step 2 by solving the SOCP problem \eqref{eq:3_42} with IPM, which requires approximately ${\mathcal{O}}\left( N_\mathrm{T}^3 L^3 \right)$.

\subsubsection{ADMM} 
This stage includes the iterative design of the radar receive filters $\{{\bf U}_k\}$, transmit waveform ${\bf W}$, BD-RIS matrices $\left( {\bm \Phi}_\mathrm{T} , {\bm \Phi}_\mathrm{R} \right)$, auxiliary variable $\left\{{\bm \Theta}_{g}\right\}$ and dual variables $\left\{ {\bm \Lambda}_{g}^n \right\}$.
Updating radar receive filters $\{{\bf U}_k\}$ requires ${\mathcal{O}}\left( K N_\mathrm{R}^3 \right)$.
Solving problem \eqref{eq:3_30} for updating ${\bf W}$ with IPM method needs complexity ${\mathcal{O}}\left( N_\mathrm{T}^3 L^3 \right)$.
The complexity of updating BD-RIS matrices $\left( {\bm \Phi}_\mathrm{T} , {\bm \Phi}_\mathrm{R} \right)$ can be upper bounded by ${\mathcal{O}}\left( M^3N_\text{S}^3 \right)$.
Using \textbf{\textit{Theorem \ref{the:1}}} to update auxiliary variable $\left\{{\bm \Theta}_{g}\right\}$ requires complexity of ${\mathcal{O}}\left( GM^3 \right)$.
Updating the dual variables $\left\{ {\bm \Lambda}_{g}^n \right\}$ using the closed-form solution in \eqref{eq:3_19e} requires the complexity of ${\mathcal{O}}\left( GM^2 \right)$.
Therefore, by disregarding lower-order complexities, the overall complexity of the ADMM framework is ${\mathcal{O}}( N_{\rm 0} (  N_\mathrm{T}^3 L^3 + M^3N_\text{S}^3  ) )$, where $N_0$ denotes the maximum number of iterations.

\begin{remark}
	When considering extremely large-scale BD-RIS scenario, i.e., $N_\textnormal{S}$$\to$$\infty$, the overall complexity becomes ${\mathcal{O}}( N_{\rm 0}  M^3N_\textnormal{S}^3 ) = {\mathcal{O}}( N_{\rm 0}  N_\textnormal{S}^6/G^3 )$.
	For CW-SC BD-RIS case ($G=N_\textnormal{S}$), overall complexity reduces to ${\mathcal{O}}( N_{\rm 0}   N_\textnormal{S}^3  )$.
	For CW-FC BD-RIS case ($M=1$), overall complexity becomes ${\mathcal{O}}( N_{\rm 0}   N_\textnormal{S}^6  )$.
	Thus, CW-FC BD-RIS requires the highest complexity, and the CW-GC BD-RIS case encounters moderate complexity ${\mathcal{O}}( N_{\rm 0}  N_\textnormal{S}^6/G^3 ) , 1< G < N_\textnormal{S}$, achieving a performance-complexity trade-off.
\end{remark}

\subsection{Convergence Analysis} 
According to the updates of $\left\{{\bf{U}}_k^n  \right\}, {{\bf{W}}^n, \left\{ {{\bf{\Phi }}_{g}^{n}} \right\} },$ and $ \left\{ {\bm \Theta}_{g}^n \right\}$ described in the previous subsections, the following propositions can be concluded.
\begin{proposition}\label{pro:1}
	Let $\left\{ \left\{ \mathbf{U}_k^n \right\} , \mathbf{W}^n, \left\{ \mathbf{\Phi }_{g}^n \right\},   \left\{ {\bm \Theta}_{g}^n \right\} \right\}$ be a sequence generated by the algorithm \ref{alg:1} and 
	assume that $\lim\limits_{n\to\infty} {\bm \Lambda}_g^{n+1} - {\bm \Lambda}_g^{n} = \mathbf{0}, \forall g$.
	Then there exists a limit point $\left\{ \left\{ \bar{\mathbf{U}}_k \right\} , \bar{\mathbf{W}}, \left\{ \bar{\mathbf{\Phi }}_g \right\},   \left\{ \bar{{\bm \Theta}}_g \right\} \right\}$, which is a stationary point and sub-optimal solution to $\mathcal{P}_{\mathrm{AL}}^2$.
\end{proposition}
\begin{IEEEproof}
	Please refer to Appendix \ref{Proof_2}.
\end{IEEEproof}

\begin{figure}[!t]
	\centering
	\includegraphics[width = 1\linewidth]{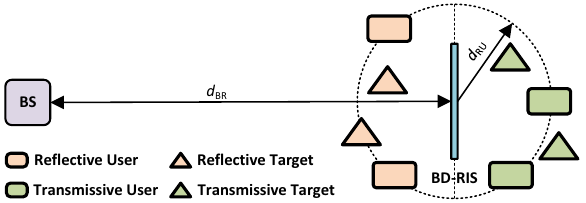}
	\caption{An illustration of the relative position among the DFBS, BD-RIS, users and targets.}
	\label{fig:schematic}
\end{figure}

\section{Performance Evaluation}\label{sec:4}
In this section, we provided extensive simulation results to validate the effectiveness of the proposed algorithm and the performance of the proposed BD-RIS aided DFRC system.

\subsection{System Setup}

We assume that the DFBS equipped with $N_\mathrm{T} = 8$ antennas transmits QPSK symbols ($\mathbb{M} = 4$) to $U = 4$ downlink users and detects $K=4$ targets with the assistance of a BD-RIS having $N_\text{S} = 16$ cells.
The radar sensing receiver colocated with the BD-RIS has $N_\mathrm{R} = 8$ receive elements.
Without loss of generality, we assume that the DFBS transmit antennas, BD-RIS antennas, and radar receiver antennas are arranged as uniform linear arrays (ULAs) with half-wavelength spacing.
The code length is $L = 16$ and the power budget at the DFBS is set as $ E = 10$ W.
The noise power at the users and radar sensing receiver are set as $\sigma_{\mathrm{c},u}^2 = \sigma_\mathrm{r}^2 = -100 \text{ dBm}, \forall u$.
The communication QoS threshold is set the same for all users, i.e., ${\Gamma_{u,l}} = {\Gamma} , \forall u,l$.
In addition, the distance-dependent path-loss is modeled as $\eta \left( d \right) = \aleph \left( {d}/{d_0} \right)^{-\ell}$, where $\aleph = -30 {\text{ dB}}$ denotes the signal attenuation at the reference distance $d_0 = 1 \text{ m}$, and $\ell$ represents the path-loss exponent.
We set the path-loss exponents for the DFBS$\to$BD-RIS, BD-RIS$\to$user, BD-RIS$\to$target, and BD-RIS$\to$clutter as 2.2, 2.2, 2, and 2, respectively.

\begin{table}[t]
	\centering
	\caption{Information of Targets and Clutters in Transmissive Area.}
	\begin{tabular}{|c|c|c|c|c|}
		\hline
		Index 		&No. 	& Range (m)     	& Azimuth ($^\circ$)    & RCS (dB) \\ \hline\hline
		Target 1   	&1  	& 10        		& 30            		& 10        \\ \hline
		Target 2   	&1 		& 19        		& -20             		& 10        \\ \hline
		Clutter 1 	& 10    & 14            	& {[}25:35{]} 			& 25       \\ \hline 
		Clutter 2 	&5    	& {[}17:21{]}		& -5      				& 25       \\ \hline
		Clutter 3 	&10    	& {[}10:14{]}  		& -50             		& 25       \\ \hline
		Clutter 4 	&5    	& 5 				& {[}-15:-25{]}  		& 25       \\ \hline
	\end{tabular}
	\label{tab_1}
\end{table}

\begin{table}[t]
	\centering
	\caption{Information of Targets and Clutters in Reflective Area.}
	\begin{tabular}{|c|c|c|c|c|}
		\hline
		Index 		&No.	& Range (m)     	& Azimuth ($^\circ$)    & RCS (dB) \\ \hline\hline
		Target 3   	&1  	& 10        		& -10           		& 10       \\ \hline
		Target 4  	&1 		& 15        		& 20           			& 10       \\ \hline
		Clutter 5 	&10    	& 10            	& {[}15:25{]} 			& 25 	\\ \hline 
		Clutter 6 	&5    	& {[}13:17{]}		& -20     				& 25       \\ \hline
		Clutter 7 	&10    	& 20  				& {[}15:25{]}       	& 25       \\ \hline
		Clutter 8 	&5    	& {[}14:18{]} 		& 50  					& 25       \\ \hline
	\end{tabular}
	\label{tab_2}
\end{table}

A two-dimensional coordinate system is shown in Fig. \ref{fig:schematic} to demonstrate the position among different devices.
The DFBS and BD-RIS are located at $\left( -20\text{ m},0\text{ m}\right)$ and $\left( 0\text{ m},0\text{ m}\right)$, respectively, which results in the distance between DFBS and BD-RIS as $d_{\rm BR} = 20\text{ m}$.
We assume $U_{\rm T} = 2$ and $U_{\rm R} = 2$ users are randomly located at transmissive and reflective areas with the same distance $d_{\rm RU} = 16\text{ m}$, respectively.
The DFBS$\to$BD-RIS and BD-RIS$\to$user channels are assumed to follow the Rician fading model with the Rician factor being 3 dB.
For the radar function, we assume $K_{\rm T} = 2$ targets and 4 groups ($Q_{\rm T} = 30$) of strong clutter sources are located in the transmissive area, whose detailed information is presented in Tables \ref{tab_1}.
Similarly, we assume $K_{\rm R} = 2$ targets and 4 groups ($Q_{\rm R} = 30$) of strong clutter sources are located in the reflective area, whose detailed information is presented in Tables \ref{tab_2}.
Moreover, we assume the range resolution as $\Delta d = 1\text{ m}$, which indicates the radar sampling rate $f_s = 150\text{ MHz}$.
Combining Table \ref{tab_1} and the path-loss model, the ratio of the propagation coefficients of the three radar targets is $\zeta _1^2:\zeta _2^2 \approx 1:0.0767$ \cite{liu2022joint,aubry2021reconfigurable,buzzi2021radar,lu2021target}, indicating that target 2 is the weakest target in transmissive area.
Following the same approach, we obtain $\zeta _3^2:\zeta _4^2 \approx 1:0.1975$, indicating that target 4 is the weakest target in the reflective area.

\subsection{Benchmark Schemes}
For comparison, we consider the following two benchmark schemes in the simulations.
\subsubsection{Benchmark 1}
The radar-only case is selected as the upper bound of the radar performance. 
We obtain this benchmark by removing the downlink users, where the resultant problem can be tackled by modifying the proposed algorithm. 
\subsubsection{Benchmark 2}
We include the STAR-RIS aided DFRC system as benchmark \cite{wang2022stars}.
As discussed in Sec. II-B \textit{Special Case 1}, the STAR-RIS is a special case of BD-RIS. 
Therefore, this benchmark can be achieved through modifying the proposed algorithm, specifically by setting $G = N_{\text{S}}$.
\subsubsection{Benchmark 3}
We consider a doulbe-RIS case where one diagonal RIS working on the reflective mode while another working on the transmissive mode are adjacently placed to achieve full-space coverage \cite{wang2022stars}.
This baseline is a special case of BD-RIS with CW-SC where  $\mathbf{\Phi}_{{\mathrm{T}}}=\text{Diag}([\phi_{{\mathrm{T}},1},\cdots,\phi_{{\mathrm{T}},\frac{N_\text{S}}{2}}],\mathbf{0}_{1\times \frac{N_\text{S}}{2}})$ and $\mathbf{\Phi}_{{\mathrm{R}}}=\text{Diag}(\mathbf{0}_{1\times \frac{N_\text{S}}{2}},[\phi_{{\mathrm{R}},1},\cdots,\phi_{{\mathrm{R}},\frac{N_\text{S}}{2}}])$.
Therefore, we can obtain this benchmark by modifying the proposed algorithm.

\subsection{Simulation Results}

\begin{figure}[!t]
	\centering
	\subfigure[]{
		\includegraphics[width = 0.85\linewidth]{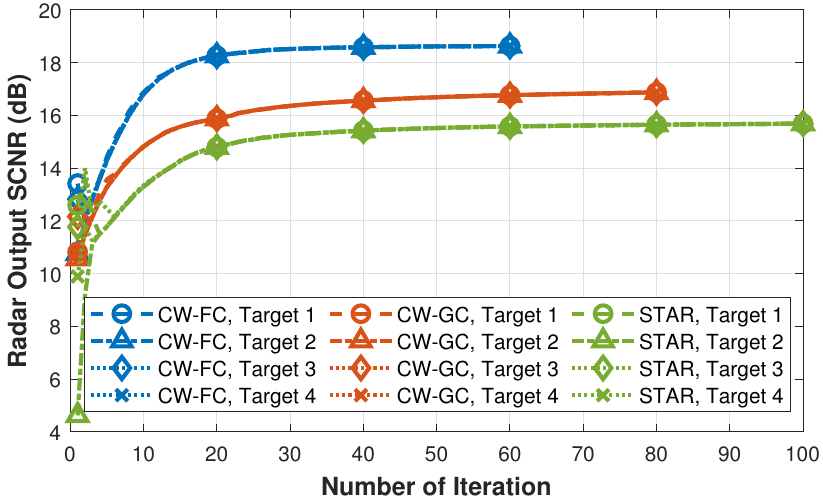}
	}
	\subfigure[]{
		\includegraphics[width = 0.85\linewidth]{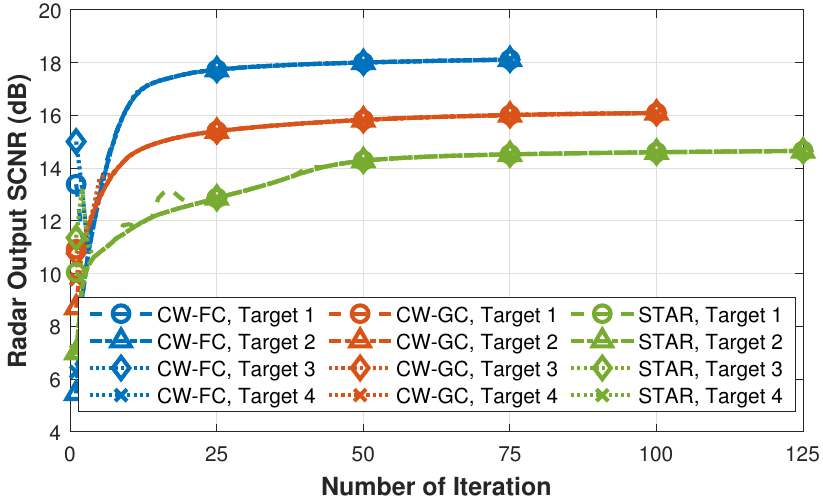}
	}
	\caption{Radar output SCNR versus the number of iterations. (a) communication threshold $\Gamma = 0{\text{ dB}}$, (b) communication threshold $\Gamma = 12{\text{ dB}}$.}
	\label{fig:conv}
\end{figure}

\subsubsection{Convergence Performance} 
In Fig. \ref{fig:conv}, we investigate the convergence of the proposed Algorithm \ref{alg:1} for different BD-RIS architectures.
It can be observed that the proposed algorithm quickly converges to a stationary point.
Specifically, after several iterations, all targets have nearly the same SCNR value, demonstrating that our algorithm can achieve fairness for multiple targets.
Moreover, the CW-FC architecture enjoys faster convergence than other architectures under the same communication threshold. 
At the same time, the STAR-RIS (CW-SC) requires nearly twice as many iterations of CW-FC to converge.
For the same architecture, the proposed algorithm with a large communication threshold $\Gamma$ needs more iterations to converge.
This is due to the fact that if the intended communication threshold $\Gamma$ is higher, fewer DoFs in the optimization problem can be used.

\begin{figure}[!t]
	\centering
	\includegraphics[width = 0.9\linewidth]{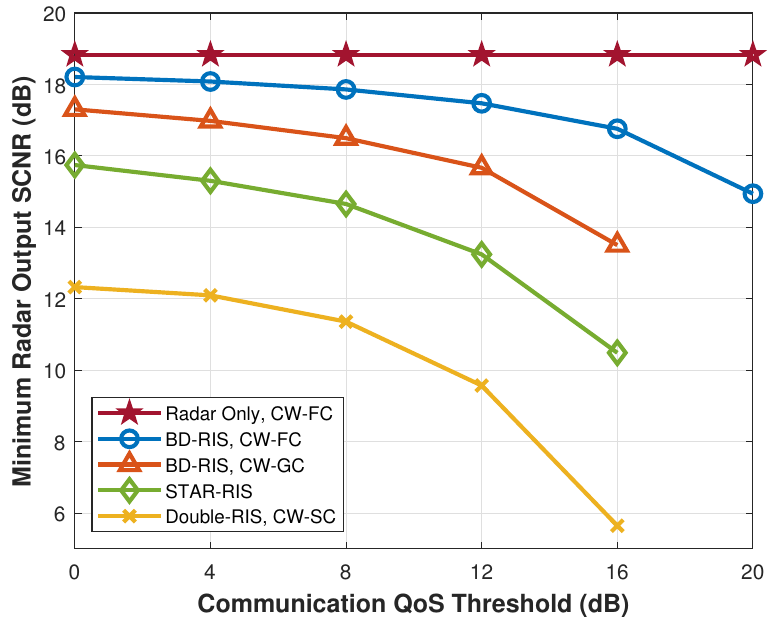}
	\caption{Minimum radar output SCNR versus the communication threshold $\Gamma$ for different architecture.}
	\label{fig:Thres_vs_SCNR}
\end{figure}

\subsubsection{System Performance with Varying Parameters}
In Fig. \ref{fig:Thres_vs_SCNR}, we study the minimum radar output SCNR versus the communication threshold $\Gamma$ for different architectures.
As expected, the radar output SCNR monotonically decreases with $\Gamma$. 
This is because when the intended $\Gamma$ is higher, less resource can be used to maximize the radar SCNR, which indicates that there is a trade-off between communication QoS and radar output SCNR.
Meanwhile, the proposed algorithm with different architectures outperform the conventional RIS, which validates the advantage of deploying BD-RIS.
In addition, the output SCNR gap between CW-FC/GC and STAR-RIS (CW-SC) becomes large with increasing communication QoS requirement, which indicates that the advantage of CW-FC/GC BD-RIS is more prominent in high communication QoS requirement scenarios.

\begin{figure}[!t]
	\centering
	\includegraphics[width = 0.9\linewidth]{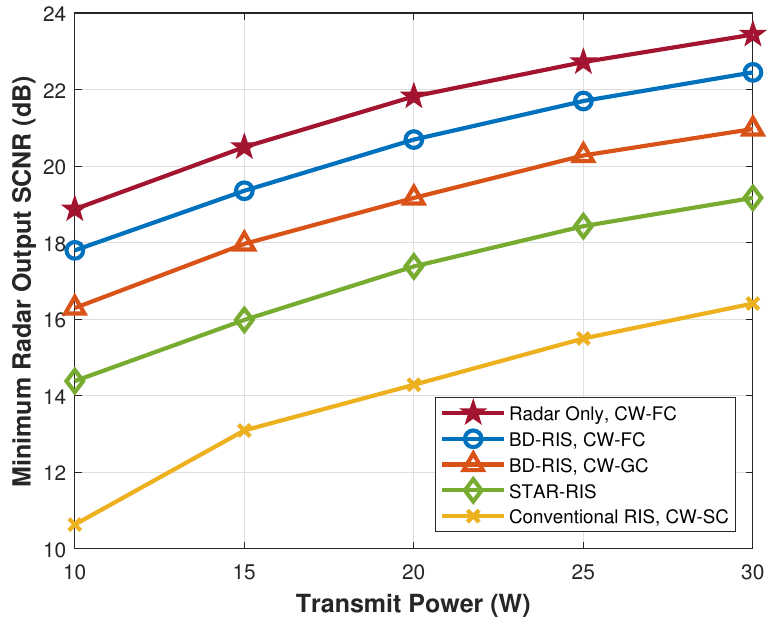}
	\caption{Minimum radar output SCNR versus the transmit power $E$ with communication threshold $\Gamma = 10 \text{ dB}$ for different architectures.}
	\label{fig:Power_vs_SCNR}
\end{figure}
Fig. \ref{fig:Power_vs_SCNR} displays the minimum radar output SCNR as a function of transmit power $E$ under different architectures.
It can be observed that the output SCNR for all schemes grows with the increase of transmit power $E$. 
Meanwhile, the growth of SCNR becomes slow when the transmit power is substantially large for all considered architectures.
This is because we can improve transmit power to boost system performance to some degree, but excessive power will not improve performance further.
By combining the findings from Figs. \ref{fig:Thres_vs_SCNR} and \ref{fig:Power_vs_SCNR}, it is evident that the BD-RIS scheme with CW-FC/SC can enhance DFRC performance.

\begin{figure}[!t]
	\centering
	\includegraphics[width = 0.9\linewidth]{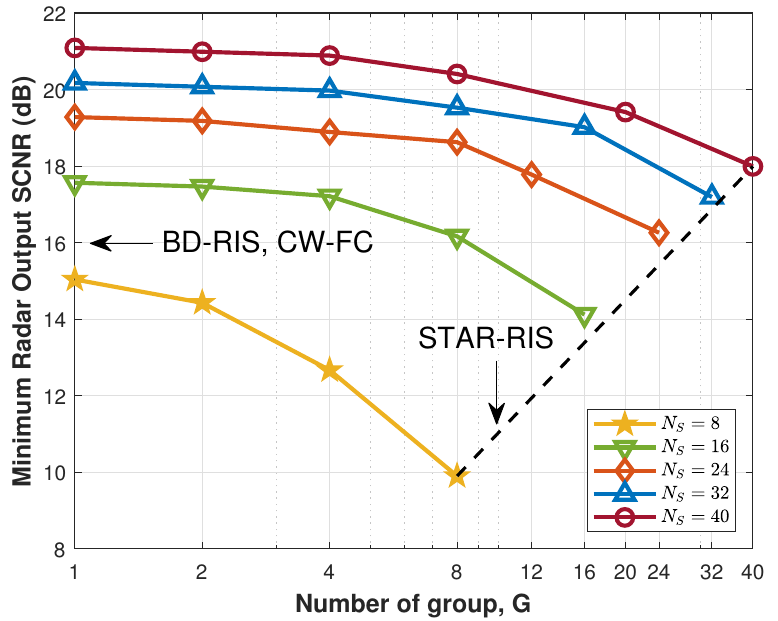}
	\caption{Minimum radar output SCNR versus the number of groups $G$ with different BD-RIS cells $N_S$ and communication threshold $\Gamma = 10 \text{ dB}$.}
	\label{fig:SINR_Cell_Ele}
\end{figure}

In Fig. \ref{fig:SINR_Cell_Ele}, we present the minimum radar SCNR versus the number of groups $G$ with different numbers of BD-RIS cells.
We observe that with the same number of groups, the radar output SCNR increases with the increasing number of BD-RIS cells.
The performance enhancement comes from the additional DoF of passive beamforming induced by the increasing number of cells, and the joint design of transmit waveform, the BD-RIS with more general constraints, and the matched filters, which also confirms the results in \cite{li2022beyondtwc}.
More importantly, the slope of each curve becomes steeper with the increasing number of groups, which indicates that the number of non-zero elements of BD-RIS matrices plays a significant role in increasing system performance.

\begin{figure}[!t]
	\centering
	\includegraphics[width = 0.9\linewidth]{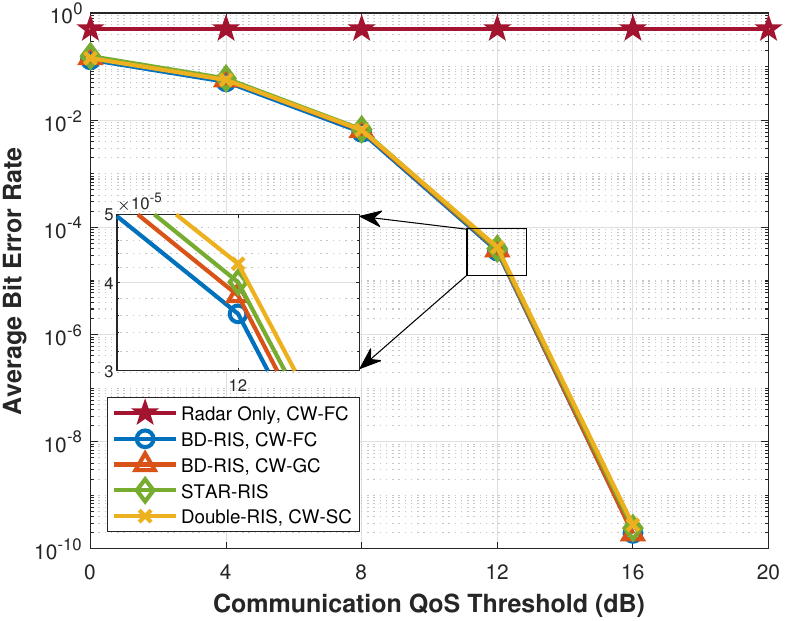}
	\caption{Average bit error rate versus the communication threshold $\Gamma$ for different architectures.}
	\label{fig:BER}
\end{figure}

\subsubsection{Communication Performance}

In Fig. \ref{fig:BER}, we plot the average BER versus the communication threshold $\Gamma$ for different architectures.
We observe that with the increase of the communication QoS threshold $\Gamma$, the BER dramatically decreases, indicating the effectiveness of SLB.
Besides, for the same QoS threshold $\Gamma$, the similar BER performance is obtained by different architectures.
This is because we set the communication QoS as constraints in problem $\mathcal{P}^1$.
Combining Figs. \ref{fig:Thres_vs_SCNR} and \ref{fig:BER}, we can conclude that by choosing an appropriate value of the QoS threshold $\Gamma$, the BD-RIS aided DFRC system can achieve satisfactory radar performance while guaranteeing high communication QoS with an extremely low BER.

\begin{figure}[!t]
	\centering
	\subfigure[]{
		\includegraphics[width = 0.85\linewidth]{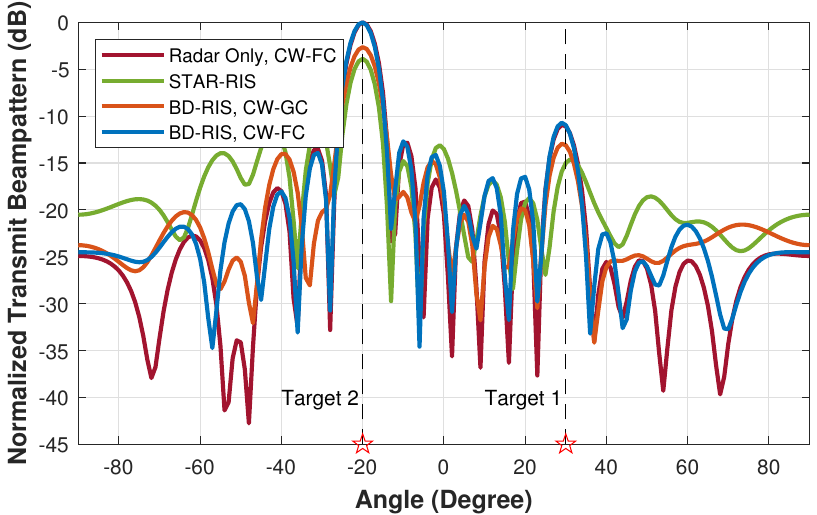}
	}
	\subfigure[]{
		\includegraphics[width = 0.85\linewidth]{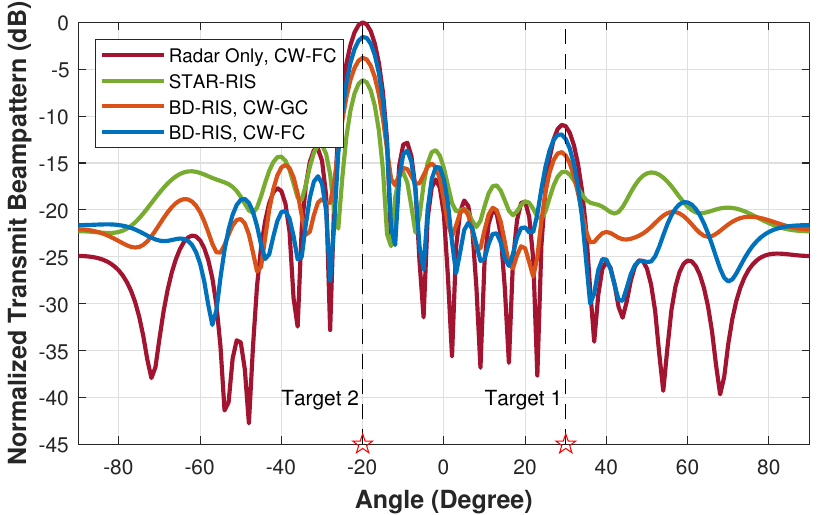}
	}
	\caption{Transmit beampattern towards the transmissive area for different architectures. (a) Communication threshold $\Gamma = 0\text{ dB} $; (b) communication threshold $\Gamma = 12\text{ dB} $.}
	\label{fig:Transmit_BP}
\end{figure}

\subsubsection{Radar Performance}
In Fig. \ref{fig:Transmit_BP}, we present the transmit beampattern towards the transmissive area obtained by the proposed algorithm.
Results show that regardless of BD-RIS architectures, the transmit power mainly concentrates around the two targets, which guarantees a high SCNR output at target directions.
Moreover, the BD-RIS with CW-FC/GC architectures can focus more energy toward targets and has a lower sidelobe than that with STAR-RIS (CW-SC) architecture, thanks to the more flexible passive beamforming control provided by the CW-FC/GC architectures.
We also observe that the transmit power towards target 2 is much higher than target 1.
This is because, as mentioned earlier, target 2 is the weakest one, which needs more energy to improve the output radar SCNR.
In addition, the transmit beampattern performance for BD-RIS with all architectures gets worse with larger communication QoS thresholds, which confirms the conclusion in  Fig. \ref{fig:Thres_vs_SCNR}.

\begin{figure}[!t]
	\centering
	\subfigure[Radar-only, CW-FC]{
		\includegraphics[width = 0.47\linewidth]{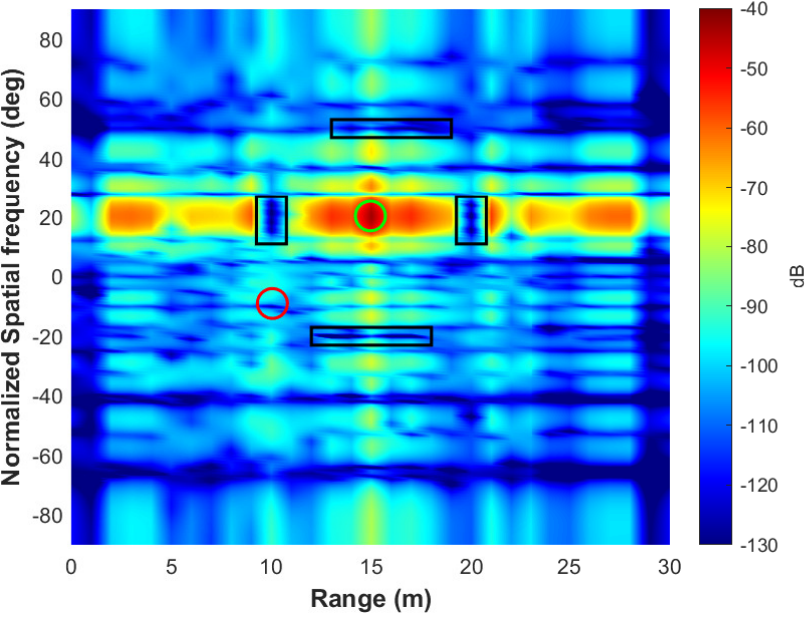}
	}
	\hspace{-1em}
	\subfigure[BD-RIS, CW-FC]{
		\includegraphics[width = 0.47\linewidth]{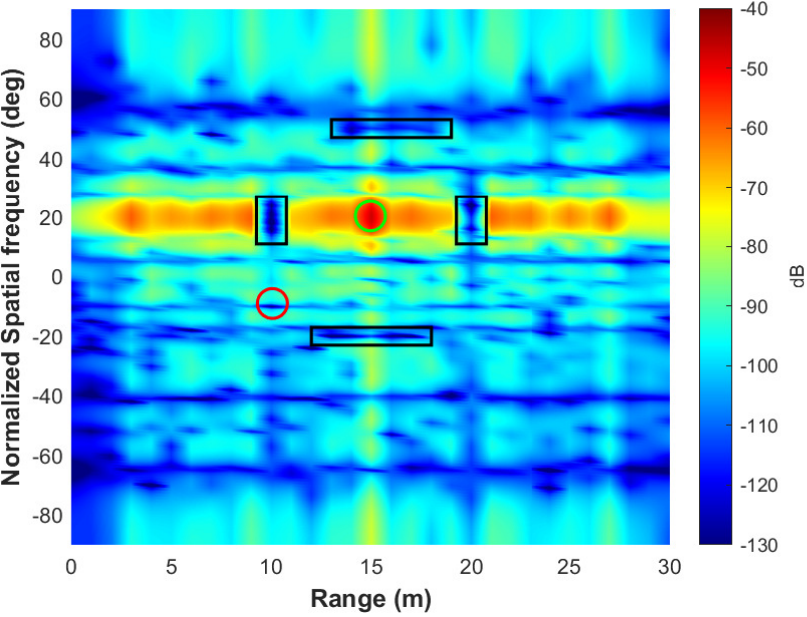}
	}
	\subfigure[BD-RIS, CW-GC]{
		\includegraphics[width = 0.47\linewidth]{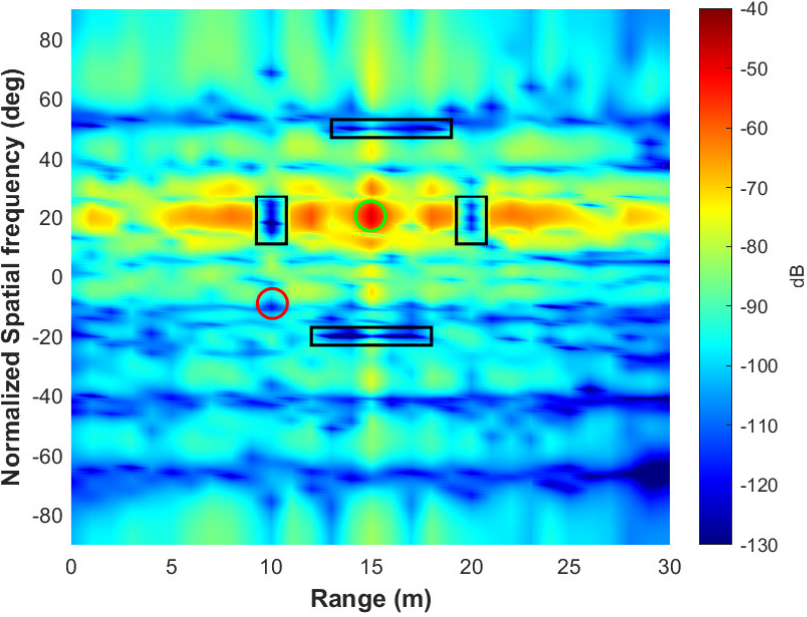}
	}
	\hspace{-1em}
	\subfigure[STAR-RIS (CW-SC)]{
		\includegraphics[width = 0.47\linewidth]{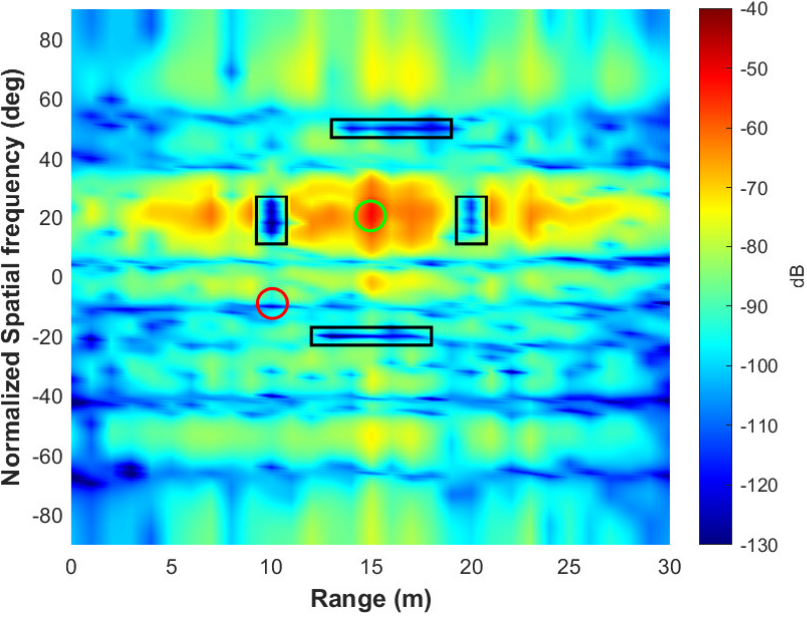}
	}
	\caption{The space-range beampattern behaviors of the receive weights for the target 4 detection with communication threshold $\Gamma = 12\text{ dB}$.}
	\label{fig:Recive_BP}
\end{figure}

Fig. \ref{fig:Recive_BP} shows the space-range beampattern of the designed waveform when BD-RIS has different architectures, where the beampattern of the $k$-th target is computed as $P_R^k\left( \theta , l \right) = | {\rm Tr} \{ \left( {\bf U}_k^\star \right)^H {\bf A}\left( \theta \right) {\bm \Phi}_\imath {\bf G} {\bf W}^\star {\bf J}_{r_l} \}|^2 , k \in \mathcal{K}_\imath , \imath \in \{{\rm T} , {\rm R}\}$ \cite{cheng2018spectrally,cui2013mimo,de2008design}.
Without loss of generality, we take target 4 ($k=4$) as an example to illustrate the space-range behavior of the designed waveform.
Results show that the space-range beampattern can form a mainlobe at the location of the target $k=4$ (green circle), but achieve null points at the locations of the other non-of-interest target (red circle) and strong clutter sources (black rectangles) for all proposed architectures.
This phenomenon can be explained as follows: 
\textit{i)} To detect target $k$, the other targets are regarded as interference.
\textit{ii)} BD-RIS with more general architectures can provide more DoFs to resist strong clutter sources.

\section{Conclusion}\label{sec:5}

This paper considers the use of BD-RIS in the DFRC system in the presence of multiple targets and strong clutter sources.
We start by reviewing the BD-RIS architectures, and deriving the communication and radar models.
Our objective is to maximize the minimum radar output SCNR subject to the constraints of communication QoS, BD-RIS matrices, and power budget.
Then, a general algorithm utilizing the ADMM approach is developed to solve the resulting complicated non-convex max-min optimization problem.
Finally, simulation results demonstrate the effectiveness of the proposed design algorithm, and the superiority of employing the BD-RIS in DFRC systems in terms of enhancing both communication and radar performance. 

Based on this initial work, there are many areas ripe for future research on BD-RIS aided DFRC. 
These include prototyping real-world BD-RIS empowered DFRC systems, developing DFRC systems powered by practical BD-RIS models such as discrete-value BD-RIS and lossy interconnections BD-RIS, designing wideband waveforms, refining scenarios for target estimation, and exploring the application of multi-sector BD-RIS in DFRC systems.

\appendices{}

\section{Proof of {Proposition \ref{pro_0}} }\label{Proof_pro_1}

Since \eqref{neq:2-13} consists of three equal equations, the proof is divided into three parts. 
In the first part, we prove \eqref{neq:2-13a}. 
Specifically, the numerator in ${\text{SCNR}}_k \left(  {\bf W} , {\bm \Phi}_\imath , {\bf U}_k \right)$ maintains the following equality
\begin{equation}
	\begin{aligned}
		&  \mathbb{E} \left\{ \left| \text{Tr}( \alpha_k \mathbf{U}_k^H \mathbf{A} ( \varphi_k ) \mathbf{\Phi}_\imath \mathbf{GW} \mathbf{J}_{r_\mathrm{T}^k} ) \right|^2 \right\}  \\
		& \mathop  = \limits^{(a)}  \mathbb{E} \left\{ \left|   \mathbf{u}_k^H \text{Vec}\{  \alpha_k \mathbf{A} ( \varphi_k ) \mathbf{\Phi}_\imath \mathbf{GW} \mathbf{J}_{r_\mathrm{T}^k} \}  \right|^2 \right\}  \\
		& \mathop  = \limits^{(b)}    \mathbf{u}_k^H \mathbb{E} \left\{ |\alpha_k|^2 \right\} \text{Vec}\{   \mathbf{A} ( \varphi_k ) \mathbf{\Phi}_\imath \mathbf{GW} \mathbf{J}_{r_\mathrm{T}^k} \} \\
		& \qquad \qquad \qquad \quad  \times \text{Vec}\{  \mathbf{A} ( \varphi_k ) \mathbf{\Phi}_\imath \mathbf{GW} \mathbf{J}_{r_\mathrm{T}^k} \}^H  \mathbf{u}\\
		& \mathop  = \limits^{(c)}  \mathbf{u}_k^H {{\bf{\Psi }}_{{\rm T},k}} \mathbf{u}_k ,
	\end{aligned}
	\nonumber
	\label{neq:A-1}%
\end{equation}
where $(a)$ holds since $\text{Tr}\{\mathbf{A}^H\mathbf{B}\} = \text{Vec}(\mathbf{A})^H \text{Vec}(\mathbf{B})$; 
$(b)$ holds since $|\mathbf{a}^H\mathbf{b}|^2 = \mathbf{a}^H\mathbf{b}\mathbf{b}^H\mathbf{a}$;
$(c)$ holds since $\mathbb{E} \left\{ |\alpha_k|^2 \right\} = \zeta _k^2$ and $\text{Vec}\{  \mathbf{A} ( \varphi_k ) \mathbf{\Phi}_\imath \mathbf{GW} \mathbf{J}_{r_\mathrm{T}^k} \} = {{{\bf{\bar M}}}_{\rm T}}\left( {k,{{\bf{\Phi }}_\imath}} \right) \mathbf{w}$ with ${{{\bf{\bar M}}}_{\rm T}}\left( {k,{{\bf{\Phi }}_\imath}} \right) = {\bf{J}}_{r_\mathrm{T}^k}^T \otimes \left( {  {\bf{A}}\left( {{\varphi _k}} \right){{\bf{\Phi }}_\imath}{\bf{G}}} \right)$.
${{\bf{\Psi }}_{{\rm T},k}}$ is defined as
\begin{equation}
	\begin{aligned}
		{{\bf{\Psi }}_{{\rm T},k}} & = \zeta _k^2 \text{Vec}\{  \mathbf{A} ( \varphi_k ) \mathbf{\Phi}_\imath \mathbf{GW} \mathbf{J}_{r_\mathrm{T}^k} \}^H \text{Vec}\{  \mathbf{A} ( \varphi_k ) \mathbf{\Phi}_\imath \mathbf{GW} \mathbf{J}_{r_\mathrm{T}^k} \}\\
		& = \zeta _k^2 {{{{\bf{\bar M}}}_{\rm T}}\left( {k,{{\bf{\Phi }}_\imath}} \right)} {\bf{w}}{{\bf{w}}^H}{\left( {{{{\bf{\bar M}}}_{\rm T}}\left( {k,{{\bf{\Phi }}_\imath}} \right)} \right)^H} .
	\end{aligned}
\end{equation}

Similarly, the denominator in ${\text{SCNR}}_k \left(  {\bf W} , {\bm \Phi}_\imath , {\bf U}_k \right)$ maintains the following equality
\begin{equation}
	\begin{aligned}
		& \mathbb{E} \Big\{ \sum_{p\in\mathcal{K}_\imath, p \ne k} { {{{ \left| {{\text{Tr}}( {{\alpha _p}{{\bf{U}}_p^H}{\bf{A}}\left( {{\varphi_p}} \right){{\bf{\Phi }}_{\imath}}{\bf{GW}}{\bf J}_{r_\mathrm{T}^p}} )} \right|}^2}} } +  \\
		& \qquad \sum_{q \in \mathcal{Q}_\imath} {{{\left| {{\text{Tr}}( {{\beta _q}{{\bf{U}}^H}{\bf{A}}\left( {{\vartheta _q}} \right){{\bf{\Phi }}_\imath}{\bf{GW}}{{\bf{J}}_{r_\mathrm{C}^q}}} )} \right|}^2}}  + \sigma_\mathrm{r}^2\left\| {\bf{U}}_k \right\|_F^2 \Big\} \\
		& = \mathbf{u}_k^H ({{\bf{\Psi }}_{{\rm C},k}} + \sigma_\mathrm{r}^2 \mathbf{I}_{N_R L}) \mathbf{u}_k ,
	\end{aligned}
	\label{neq:A-2}
\end{equation}
where
\begin{equation}
	\begin{aligned}
		& {{\bf{\Psi }}_{{\rm C},k}} = \sum\limits_{p \in \mathcal{K}_\imath ,p \ne k} {\zeta _p^2 {{{{\bf{\bar M}}}_{\rm T}}\left( {p,{{\bf{\Phi }}_\imath}} \right)} {\bf{w}}{{\bf{w}}^H}{{\left( {{{{\bf{\bar M}}}_{\rm T}}\left( {p,{{\bf{\Phi }}_\imath}} \right)} \right)}^H}} \\
		& \qquad \quad + \sum\limits_{q \in \mathcal{Q}_\imath} {\xi _q^2 {{{{\bf{\bar M}}}_{\rm C}}\left( {q,{{\bf{\Phi }}_\imath}} \right)} {\bf{w}}{{\bf{w}}^H}{{\left( {{{{\bf{\bar M}}}_{\rm C}}\left( {q,{{\bf{\Phi }}_\imath}} \right)} \right)}^H}}  ,
	\end{aligned}
\end{equation}
with ${{{\bf{\bar M}}}_{\rm C}}( {q,{{\bf{\Phi }}_\imath}} ) = {\bf{J}}_{r_\mathrm{C}^q}^T \otimes \left( {  {\bf{A}}\left( {{\vartheta _q}} \right){{\bf{\Phi }}_\imath}{\bf{G}}} \right)$.

Combining \eqref{neq:A-1} and \eqref{neq:A-2}, we have 
\begin{equation}
	{\text{SCNR}}_k\left( {{\bf{W}},{{\bf{\Phi }}_\imath},{{\bf{U}}_k}} \right) = \frac{{{{\bf{u}}_k^H}{{\bf{\Psi }}_{\mathrm{T},k}}{\bf{u}}_k}}{{{{\bf{u}}_k^H} \left({{\bf{\Psi }}_{\mathrm{C},k}} + \sigma_\mathrm{r}^2{{\bf{I}}_{{N_\mathrm{R}}L}} \right) {\bf{u}}_k}}.
\end{equation}

Following the same procedure, we can also prove the second and third parts of \eqref{neq:2-13} as follows
\begin{subequations}
	\begin{align}
		{\text{SCNR}}_k\left( {{\bf{W}},{{\bf{\Phi }}_\imath},{{\bf{U}}_k}} \right) & = \frac{{{{\bf{w}}^H}{{\bm \Upsilon} _{\mathrm{T},k}} {\bf{w}}}}{{{{\bf{w}}^H}{{\bm \Upsilon} _{\mathrm{C},k}} {\bf{w}} + \sigma_\mathrm{r}^2\left\| {\bf{U}}_k \right\|_F^2}} , \\
		& = \frac{{{{\bm \phi}}_\imath^H{{\bm{\Xi }}_{{\rm T},k}}{{{\bm \phi}}_\imath}}}{{{{\bm \phi}}_\imath^H{{\bm{\Xi }}_{\mathrm{C},k}}{{{\bm \phi}}_\imath} + \sigma_\mathrm{r}^2\left\| {\bf{U}}_k \right\|_F^2}} , 
	\end{align}
\end{subequations}
where
\begin{equation}
	\begin{aligned}
		{{\bm \Upsilon} _{{\rm T},k}} = &  \zeta _k^2{\left( {{{{\bf{\bar M}}}_{\rm T}}\left( {k,{{\bf{\Phi }}_\imath}} \right)} \right)^H}{\bf{u}}_k{{\bf{u}}_k^H} {{{{\bf{\bar M}}}_{\rm T}}\left( {k,{{\bf{\Phi }}_\imath}} \right)}  , \\
		{{\bm \Upsilon} _{{\rm C},k}} = & \sum\limits_{p \in \mathcal{K}_\imath,p \ne k} {\zeta _p^2{{\left( {{{{\bf{\bar M}}}_{\rm T}}\left( {p,{{\bf{\Phi }}_\imath}} \right)} \right)}^H}{\bf{u}}_k{{\bf{u}}_k^H} {{{{\bf{\bar M}}}_{\rm T}}\left( {p,{{\bf{\Phi }}_\imath}} \right)} } \\
		& + \sum\limits_{q \in \mathcal{Q}_\imath} {\xi _q^2{{\left( {{{{\bf{\bar M}}}_{\rm C}}\left( {q,{{\bf{\Phi }}_\imath}} \right)} \right)}^H}{\bf{u}}_k{{\bf{u}}_k^H} {{{{\bf{\bar M}}}_{\rm C}}\left( {q,{{\bf{\Phi }}_\imath}} \right)} }  , \\
		{{\bf{\Xi }}_{{\rm T},k}} = & \zeta _k^2{\left( {{{{\bf{\tilde M}}}_{\rm T}}\left( {k,{\bf{W}}} \right)} \right)^H}{\bf{u}}_k{{\bf{u}}_k^H} {{{{\bf{\tilde M}}}_{\rm T}}\left( {k,{\bf{W}}} \right)}  , \\
		{{\bf{\Xi }}_{{\rm C},k}} = & \sum\limits_{p \in \mathcal{K}_\imath, p \ne k} {\zeta _p^2{{\left( {{{{\bf{\tilde M}}}_{\rm T}}\left( {p,{\bf{W}}} \right)} \right)}^H}{\bf{u}}_k{{\bf{u}}_k^H} {{{{\bf{\tilde M}}}_{\rm T}}\left( {p,{\bf{W}}} \right)} } \\
		& + \sum\limits_{q \in \mathcal{Q}_\imath} {\xi _q^2{{\left( {{{{\bf{\tilde M}}}_{\rm C}}\left( {q,{\bf{W}}} \right)} \right)}^H}{\bf{u}}_k{{\bf{u}}_k^H} {{{{\bf{\tilde M}}}_{\rm C}}\left( {q,{\bf{W}}} \right)} } ,
	\end{aligned}
	\nonumber
	\label{eq:3_14}%
\end{equation}
with
${{{\bf{\tilde M}}}_{\rm T}}( {k,{\bf{W}}} ) = ( {{\bf{J}}_{r_\mathrm{T}^k}^T{{\bf{W}}^T}{{\bf{G}}^T}} )  \otimes {\bf{A}}( {{\varphi _k}} )$,
${{{\bf{\tilde M}}}_{\rm C}}( {q,{\bf{W}}} ) = ( {{\bf{J}}_{r_\mathrm{C}^q}^T{{\bf{W}}^T}{{\bf{G}}^T}} )  \otimes {\bf{A}}( {{\vartheta _k}} )$.

The proof is completed.

\section{Proof of {Theorem \ref{nthe:1}} }\label{Proof_nthe_1}

Problem $\mathcal{P}^2_{{\rm AL},{\bf W}}$ is hard to settle due to the non-smooth objective function and complicated non-convex constraints.
To simplify the design, we first equivalently transform the objective into a smooth form by introducing an auxiliary variable $\gamma$, which yields the following problem
\begin{subequations}
	\begin{numcases}{\mathcal{P}^{2-1}_{{\rm AL},{\bf W}}}
		\mathop {\max }\limits_{{\bf{W}} , \gamma}  \;\; \gamma \\
		\;\; {\rm {s.t.}}  \;\; \min\limits_{\forall k} \frac{{{{\bf{w}}^H}{{\bm \Upsilon} _{{\rm T},k}} {\bf{w}}}}{{{{\bf{w}}^H}{{\bm \Upsilon} _{{\rm C},k}} {\bf{w}} + \sigma_\mathrm{r}^2\left\| {\bf{U}}_k \right\|_F^2}} \ge \gamma , \label{eq:3_24b}\\
		\qquad\;\; \gamma \ge 0 , \\
		\qquad\;\; \eqref{eq:3_23b}, \eqref{eq:3_23c}. \label{eq:3_24d}
	\end{numcases}
	\label{eq:3_24}%
\end{subequations}
Then, we deal with constraints \eqref{eq:3_24b}, and \eqref{eq:3_23b} step-by-step detailed as follows.

\textit{Step 1: Successive Convex Approximation (SCA) to \eqref{eq:3_24b}.} 
We first rewrite constraint \eqref{eq:3_24b} as
\begin{equation}
	{\bf{w}}^H{{\bm \Upsilon} _{{\rm C},k}} {\bf{w}} - \frac{ {{{\bf{w}}^H}{{\bm \Upsilon} _{{\rm T},k}} {\bf{w}}} }{ \gamma } + \sigma_\mathrm{r}^2\left\| {\bf{U}}_k \right\|_F^2 \le 0, \forall k,
	\label{eq:3_25}
\end{equation}
where the second term is a composite function with both $\mathbf{w}$ and $\gamma$, which makes \eqref{eq:3_25} non-convex and hard to tackle.
To facilitate the joint design of problem \eqref{eq:3_24}, we perform SCA to \eqref{eq:3_25} and propose the following lemma.
\begin{lemma}\label{lem_1}
	Define $f\left( {\bf w} , \gamma \right) = \frac{{\bf w}^H{\bm \Upsilon}{\bf w}}{\gamma}$.
	If ${\bm \Upsilon}$ is positive definite and $\gamma > 0$, we have
	\begin{enumerate}[leftmargin=13pt]
		\item $f\left( {\bf w} , \gamma \right) = \frac{{\bf w}^H{\bm \Upsilon}{\bf w}}{\gamma}$ is jointly convex on $\mathbf{w}$ and $\gamma$.
		
		\item A minorizer of $f\left( {\bf w} , \gamma \right) = \frac{{\bf w}^H{\bm \Upsilon}{\bf w}}{\gamma}$ is 
		\begin{equation}
			f\left( {{\bf{w}},\gamma;{{\bf{w}}^n},{\gamma^n}} \right) = \frac{{2\Re \left\{ {({\bf{w}}^n)^H{{\bm \Upsilon}\bf{w}}} \right\}}}{{{\gamma^n}}} - \gamma\frac{{({\bf{w}}^n)^H{\bm{\Upsilon}}{{\bf{w}}^n}}}{{(\gamma^n)^2}} ,
			\notag
		\end{equation} 
		where $(\mathbf{w}^n , \gamma_n)$ is the point at $n$-th iteration.
	\end{enumerate}
\end{lemma}
\begin{IEEEproof}
	Please refer to Appendix \ref{Proof_lem_1}.
\end{IEEEproof}
Using \textit{\textbf{Lemma \ref{lem_1}}}, we conduct the SCA on constraint \eqref{eq:3_25} at point $\left( {\bf w}^n , \gamma^n \right)$.
Thus, we can safely approximate \eqref{eq:3_25} as
\begin{equation}
	\begin{aligned}
		{\bf{w}}^H{{\bm \Upsilon} _{{\rm C},k}} {\bf{w}} & - \frac{ 2 \Re \left\{ {{({\bf{w}}^n)^H}{{\bm \Upsilon} _{{\rm T},k}} {\bf{w}}} \right\} }{ \gamma^n } \\
		& + \gamma \frac{   {{({\bf{w}}^n)^H}{{\bm \Upsilon} _{{\rm T},k}} {\bf{w}}^n} }{ (\gamma^n)^2 }  + \sigma_\mathrm{r}^2\left\| {\bf{U}}_k \right\|_F^2 \le 0, \forall k,
	\end{aligned}
	\label{eq:3_27}
\end{equation}
where $\gamma^n$ is computed by 
\begin{equation}
	\gamma^n = \min\limits_{\forall k} \frac{{{({\bf{w}}^n)^H}{{\bm \Upsilon} _{{\rm T},k}} {\bf{w}}^n}}{{{({\bf{w}}^n)^H}{{\bm \Upsilon} _{{\rm C},k}} {\bf{w}}^n + \sigma_\mathrm{r}^2\left\| {\bf{U}}_k \right\|_F^2}} .
\end{equation}

\textit{Step 2: Reformulation to \eqref{eq:3_23b}.} 
After some algebraic manipulations, we rewrite \eqref{eq:3_23b} as 
\begin{subequations}
	\begin{numcases}{\eqref{eq:3_23b} \Leftrightarrow}
		\Re \left\{ {{\bf{\bar h}}_{u,1}^H\left( {{{\bf{\Phi }}_\imath}} \right){\bf{w}}\left[ l \right]} \right\} \ge \sqrt {\sigma _{\mathrm{c},u}^2{\Gamma _{u,l}}} \sin \Omega, \label{eq:3_28a} \\
		\Re \left\{ {{\bf{\bar h}}_{u,2}^H\left( {{{\bf{\Phi }}_\imath}} \right){\bf{w}}\left[ l \right]} \right\} \ge \sqrt {\sigma _{\mathrm{c},u}^2{\Gamma _{u,l}}} \sin \Omega,  \label{eq:3_28b}
	\end{numcases}
	\label{eq:3_28}%
\end{subequations}
where ${\bf{\bar h}}_{u,1}\left( {{{\bf{\Phi }}_\imath}} \right) = {{\bf{G}}^H}{\bf{\Phi }}_\imath^H{{\bf{h}}_u} ( \sin \Omega  + {e^{\jmath \frac{\pi }{2}}}\cos \Omega  ) e^{ - \jmath \angle \left( {{{\bf{s}}_u}\left[ l \right]} \right)}$ and ${\bf{\bar h}}_{u,2}\left( {{{\bf{\Phi }}_\imath}} \right) = {{\bf{G}}^H}{\bf{\Phi }}_\imath^H{{\bf{h}}_u} ( \sin \Omega  - {e^{\jmath \frac{\pi }{2}}}\cos \Omega ) e^{ - \jmath \angle \left( {{{\bf{s}}_u}\left[ l \right]} \right)}$.

Replacing non-convex contraints in \eqref{eq:3_24} with \eqref{eq:3_27} and \eqref{eq:3_28} based on Steps 1-2, we minimize the following problem
\begin{subequations}
	\begin{numcases}{\mathcal{P}^{2-2}_{{\rm AL},{\bf W}}}
		\mathop {\min }\limits_{{\bf{W}} , \gamma}  \;\; -\gamma \\
		\;\; {\rm {s.t.}}  \;\;  \gamma \ge 0, \;\; \eqref{eq:3_27}, \eqref{eq:3_28a}, \eqref{eq:3_28b}, \eqref{eq:3_23c}.
	\end{numcases}
	\label{eq:3_30}%
\end{subequations}
Problem $\mathcal{P}^{2-2}_{{\rm AL},{\bf W}}$ is a convex SOCP problem and can be globally solved by the IPM.

The proof is completed.

\section{Proof of {Lemma \ref{lem_1}} }\label{Proof_lem_1}
First, we show the first part of \textit{\textbf{Lemma \ref{lem_1}}} as follows.
The function $g(\mathbf{x},\mathbf{Y}) = \mathbf{x}^H \mathbf{Y}^{-1} \mathbf{x}$ with $\mathbf{Y} \succ 0$ is jointly convex in $\mathbf{x}$ and $\mathbf{Y}$ \cite[Section 4]{boyd2004convex}.
Let $\mathbf{x} = {\bm \Upsilon}^{\frac{1}{2}} \mathbf{w}$ and $\mathbf{Y} = \text{Diag}(\gamma , \cdots , \gamma) \succ {\bf 0}$ with ${\bm \Upsilon} \succeq {\bf 0} $ and $\gamma > 0$, both of which are affine transformations of $\mathbf{w}$ and $\gamma$.
Thus, $f\left( {\bf w} , \gamma \right) = \frac{{\bf w}^H{\bm \Upsilon}{\bf w}}{\gamma}$ is jointly convex in ${\bf w}$ and $\gamma$ when ${\bm \Upsilon} \succeq {\bf 0} $ and $\gamma > 0$.

Next, we prove the second part of \textit{\textbf{Lemma \ref{lem_1}}} as follows.
Since $f\left( {\bf w} , \gamma \right) = \frac{{\bf w}^H{\bm \Upsilon}{\bf w}}{\gamma}$ is jointly convex in ${\bf w}$ and $\gamma$ when ${\bm \Upsilon} \succeq {\bf 0} $ and $\gamma > 0$, the first order approximation of $f\left( {\bf w} , \gamma\right)$, denoted by $f\left( {{\bf{w}},\gamma;{{\bf{w}}^n},{\gamma^n}} \right)$, is a minorizer of $f\left( {\bf w} , \gamma\right)$ at the point $\left( {\bf w}^n , \gamma^n\right)$, which is 
\begin{align}
	& f\left( {{\bf{w}},\gamma;{{\bf{w}}^n},{\gamma^n}} \right) \nonumber\\
	& = f\left( {\bf w}^n , \gamma^n\right)  + {( {\frac{{\partial f}}{{\partial {\bf{w}}}}{|_{{\bf{w}} = {{\bf{w}}^n}}}} )^T}\left( {{\bf{w}} - {{\bf{w}}^n}} \right)  \nonumber \\
	& \;\;\; + {( {\frac{{\partial f}}{{\partial {{\bf{w}}^*}}}{|_{{\bf{w}} = ({\bf{w}}^n)^*}}} )^T}\left( {{\bf{w}} - ({\bf{w}}^n)^*} \right) \nonumber \\
	&\;\;\; + {( {\frac{{\partial f}}{{\partial \gamma}}{|_{\gamma = {\gamma^n}}}} )^T}\left( {\gamma - {\gamma^n}} \right) + {( {\frac{{\partial f}}{{\partial {\gamma^*}}}{|_{\gamma = (\gamma^n)^*}}} )^T}\left( {\gamma - (\gamma^n)^*} \right) \nonumber \\
	&= \frac{{({\bf{w}}^n)^H{\bm{\Upsilon}}{{\bf{w}}^n}}}{{{\gamma^n}}} + 2\Re \left\{ \left[
	\begin{array}{l}
		\frac{{2{\bm{\Upsilon}}{{\bf{w}}^n}}}{{{\gamma_n}}} \\
		\frac{{({\bf{w}}^n)^H{\bm{\Upsilon}}{{\bf{w}}^n}}}{{(\gamma^n)^2}}
	\end{array}\right]^H \left[
	\begin{array}{c}
		{\bf w} - {\bf w}^n  \\
		\gamma - \gamma^n
	\end{array}\right]\right\}  \nonumber \\
	&= \frac{{2\Re \left\{ {({\bf{w}}^n)^H{\bf{Aw}}} \right\}}}{{{\gamma^n}}} - \gamma\frac{{({\bf{w}}^n)^H{\bm{\Upsilon}}{{\bf{w}}^n}}}{{(\gamma^n)^2}}. \nonumber
\end{align}

The proof is thereby completed.

\section{Proof of {Theorem \ref{nthe:2}} }\label{Proof_nthe_2}

Problem ${\mathcal{P}^2_{{\text{AL}}, {\left\{ {\bm \Phi}_{g} \right\}}}}$ is hard to settle due to the non-smooth objective function and complicated constraints.
We propose to solve it by using the following two steps.

\textit{Step 1: De-diagonalization.}
In this step, we handle constraints \eqref{eq:3_31c} and \eqref{eq:3_31d} by extracting the non-zero parts of matrices $\mathbf{\Phi}_\mathrm{T}$ and $\mathbf{\Phi}_\mathrm{R}$. 
Specifically, by defining ${{{\bf{\tilde \Phi }}}_\imath} = [ {{{\bf{\Phi }}_{\imath,1}}, \cdots ,{{\bf{\Phi }}_{\imath,G}}} ] , \imath \in \{ {\rm T} , {\rm R}\}$, ${\bm {\tilde \phi}_\imath} = {\rm Vec}( {\bm {\tilde \Phi}_\imath} )$ and using property of vectorization \cite{zhang2017matrix}, the first term in objective function \eqref{eq:3_31a} can be reformulated as
\begin{equation}
	\frac{{{{\bm \phi}} _\imath^H{{\bm{\Xi }}_{{\rm T},k}}{{{\bm \phi}}_\imath}}}{{{{\bm \phi}}_\imath^H{{\bm{\Xi }}_{{\rm C},k}}{{{\bm \phi}}_\imath} + \sigma_\mathrm{r}^2\left\| {\bf{U}}_k \right\|_F^2}} 
	\Leftrightarrow
	\frac{{{{\bm {\tilde \phi}}} _\imath^H{{\bm{\tilde \Xi }}_{{\rm T},k}}{{{\bm {\tilde \phi}}} _\imath}}}{{{{\bm {\tilde \phi}}} _\imath^H{{\bm{\tilde \Xi }}_{{\rm C},k}}{{{\bm {\tilde \phi}}} _\imath} + \sigma_\mathrm{r}^2\left\| {\bf{U}}_k \right\|_F^2}} ,
\end{equation}
where 
${{\bm{\tilde \Xi }}_{{\rm T},k}} = {\bf K}_G {{\bm{ \Xi }}_{{\rm T},k}} {\bf K}_G^H$ and ${{\bm{\tilde \Xi }}_{{\rm C},k}} = {\bf K}_G{{\bm{ \Xi }}_{{\rm C},k}}{\bf K}_G^H$, with 
${\bf K}_G = \text{BlkDiag} ( [ {\bf{I}}_M \otimes [ {{\bf{0}}_{M,({g - 1})M}} , {{\bf{I}}_M} , {\bf{0}}_{M,\left( {G - g} \right)M} ]]_{g = 1}^G) \in\{ 0, 1\}^{M N_\text{S} \times N_\text{S}^2}$ denoting the linear mapping matrix.

Using the basic property of matrix transformation, the last two items in objective function \eqref{eq:3_31a} can be reformulated as
\begin{equation}
	\begin{aligned}
		& \sum\limits_{g=1}^{G}{ {\Re \left\{ {\rm Tr} \left( {\bm \Lambda}_g^H \left( {\bm \Phi}_{g} - {\bm \Theta}_{g} \right) \right) \right\}} } + \frac{\varrho}{2} \sum\limits_{g=1}^{G}{  \left\| {{\bm \Phi}_{g} - {\bm \Theta}_{g} } \right\|_F^2 } \\
		\Leftrightarrow  & \sum\limits_{\imath \in \{{\rm T} , {\rm R}\}}{ \Big\{ {\Re \left\{ {\rm Tr} \left( {\bm {\tilde \Lambda}}_\imath^H \left( {\bm {\tilde \Phi}}_\imath - {\bm {\tilde \Theta}}_\imath \right) \right) \right\}}   + \frac{\varrho}{2}    \left\| {{\bm {\tilde \Phi}}_\imath - {\bm {\tilde \Theta}}_\imath } \right\|_F^2 \Big\} } \\
		& = \mathbb{P}( \{ {\bm {\tilde \Phi}}_\imath \} , \{ {\bm {\tilde \Theta}}_\imath \} , \{ {\bm {\tilde \Lambda}}_\imath \} ) ,
	\end{aligned}
	\nonumber
\end{equation}
where ${{{\bf{\tilde \Theta }}}_\imath} = [ {{{\bf{\Theta }}_{\imath,1}}, \cdots ,{{\bf{\Theta }}_{\imath,G}}} ]$, ${{{\bm{\tilde \Lambda }}}_\imath} = [ {{{\bm{\Lambda }}_{\imath,1}^H}, \cdots ,{{\bm{\Lambda }}_{\imath,G}^H}} ]^H$. 
$\bm{\Lambda}_{\mathrm{T},g}$ and $\bm{\Lambda}_{\mathrm{R},g}$ are extracted from the first $M$ rows and last $M$ rows of ${\bm \Lambda}_g$, respectively.

To de-diagonalize constraints \eqref{eq:3_31b}, we partition ${{{\bf{\bar H}}}_{u,l}}$ as
\begin{equation}
	{{{\bf{\bar H}}}_{u,l}} = \left[\begin{array}{ccc}
		{{\bf{\bar H}}}_{u,l}^{11} & \cdots & {{\bf{\bar H}}}_{u,l}^{1G} \\
		\vdots  &  \ddots  & \vdots \\
		{{\bf{\bar H}}}_{u,l}^{G1} & \cdots & {{\bf{\bar H}}}_{u,l}^{GG} \\
	\end{array}\right], \forall u, l ,
\end{equation}
where ${{\bf{\bar H}}}_{u,l}^{ij} \in {\mathbb{C}}^{M \times M}$.
By defining ${{{\bf{\tilde H}}}_{u,l}} = [ {{\bf{\bar H}}_{u,l}^{11}, \cdots ,{\bf{\bar H}}_{u,l}^{GG}} ]$ and adopting basic properties of diagonalization \cite{zhang2017matrix}, constraints \eqref{eq:3_31b} can be simplified as
\begin{subequations}
	\begin{align}
		& \frac{{\left| {\Im \left\{ {{\text{Tr}}\left\{ {{{{\bf{\bar H}}}_{u,l}}{{\bf{\Phi }}_\imath}} \right\}} \right\}} \right|}}{{\Re \left\{ {{\text{Tr}}\left\{ {{{{\bf{\bar H}}}_{u,l}}{{\bf{\Phi }}_\imath}} \right\}} \right\} - \sqrt {\sigma _{\mathrm{c},u}^2{\Gamma _{u,l}}} }} \le \tan \Omega,  \forall u,l   \\
		\Leftrightarrow & 
		\frac{{\left| {\Im \left\{ {{\text{Tr}}\left\{ {{{{\bf{\tilde H}}}_{u,l}}{{{\bf{\tilde \Phi }}}_\imath}} \right\}} \right\}} \right|}}{{\Re \left\{ {{\text{Tr}}\left\{ {{{{\bf{\tilde H}}}_{u,l}}{{{\bf{\tilde \Phi }}}_\imath}} \right\}} \right\} - \sqrt {\sigma _{\mathrm{c},u}^2{\Gamma _{u,l}}} }} \le \tan \Omega  , \forall u,l , \label{Neq:3_33b}
	\end{align}
\end{subequations}

Based on the above de-diagonalization, problem ${\mathcal{P}^2_{{\text{AL}}, {\left\{ {\bm \Phi}_{g} \right\}}}}$ can be equivalently reformulated as
\begin{subequations}
	\begin{numcases}{\mathcal{P}^{2-1}_{{\text{AL}}, {\left\{ {\bm {\tilde\Phi}}_\imath \right\}}}}
		\mathop {\min }\limits_{ \{{\bf{{\tilde\Phi} }}_\imath\} }  - \Big\{ \min\limits_{\forall k} \frac{{{{\bm {\tilde \phi}}} _\imath^H{{\bm{\tilde \Xi }}_{{\rm T},k}}{{{\bm {\tilde \phi}}} _\imath}}}{{{{\bm {\tilde \phi}}} _\imath^H{{\bm{\tilde \Xi }}_{{\rm C},k}}{{{\bm {\tilde \phi}}} _\imath} + \sigma_\mathrm{r}^2\left\| {\bf{U}}_k \right\|_F^2}}   \Big\} \notag \\
		\qquad\;\; + \mathbb{P}( \{ {\bm {\tilde \Phi}}_\imath \} , \{ {\bm {\tilde \Theta}}_\imath \} , \{ {\bm {\tilde \Lambda}}_\imath \} ) \\
		{\rm {s.t.}}  \;\; \frac{{\left| {\Im \left\{ {{\text{Tr}}\left\{ {{{{\bf{\tilde H}}}_{u,l}}{{{\bf{\tilde \Phi }}}_\imath}} \right\}} \right\}} \right|}}{{\Re \left\{ {{\text{Tr}}\left\{ {{{{\bf{\tilde H}}}_{u,l}}{{{\bf{\tilde \Phi }}}_\imath}} \right\}} \right\} - \sqrt {\sigma _{\mathrm{c},u}^2{\Gamma _{u,l}}} }} \nonumber \\
		\qquad \qquad \quad \le \tan \Omega , \forall u \in \mathcal{U}_\imath , \imath \in \{{\rm T} , {\rm R}\}. \label{Neq:3_34b}
	\end{numcases}
	\label{Neq:3_34}%
\end{subequations}
Then, we will solve problem ${\mathcal{P}^{2-1}_{{\text{AL}}, {\left\{ {\bm \Phi}_\imath \right\}}}}$ in step 2.

\textit{Step 2: Solution to Problem ${\mathcal{P}^{2-1}_{{\text{AL}}, {\left\{ {\bm \Phi}_\imath \right\}}}}$.}
Note that problem ${\mathcal{P}^{2-1}_{{\text{AL}}, {\left\{ {\bm \Phi}_\imath \right\}}}}$ has the same form as problem $\mathcal{P}^2_{{\rm AL},{\bf W}}$.
Therefore, by introducing auxiliary variable $\eta$, performing \textbf{\textit{Lemma \ref{lem_1}}} and rewriting \eqref{Neq:3_34b}, the problem ${\mathcal{P}^{2-1}_{{\text{AL}}, {\left\{ {\bm \Phi}_\imath \right\}}}}$ can be recast as
\begin{subequations}
	\begin{numcases}{\mathcal{P}^{2-2}_{{\text{AL}}, {\left\{ {\bm {\tilde\Phi}}_\imath \right\}}}}
		\mathop {\min }\limits_{ \{{\bf{{\tilde\Phi} }}_\imath\} }  - \eta + \mathbb{P}( \{ {\bm {\tilde \Phi}}_\imath \} , \{ {\bm {\tilde \Theta}}_\imath \} , \{ {\bm {\tilde \Lambda}}_\imath \} ) \\
		{\rm {s.t.}}  \; {{\bm {\tilde \phi}}} _\imath^H{{\bm{\tilde \Xi }}_{{\rm C},k}}{{{\bm {\tilde \phi}}} _\imath}  - \frac{2\Re\left\{{({{\bm {\tilde \phi}}} _{\imath}^n)^H{{\bm{\tilde \Xi }}_{{\rm T},k}}{{{\bm {\tilde \phi}}} _\imath}}\right\}}{\eta^n}  \notag  \\
		\qquad \qquad + \eta \frac{2\Re\left\{{({{\bm {\tilde \phi}}} _{\imath}^n)^H{{\bm{\tilde \Xi }}_{{\rm T},k}}{{{\bm {\tilde \phi}}} _{\imath}^n}} \right\}}{(\eta^n)^2}  \notag \\
		\qquad \qquad + \sigma_\mathrm{r}^2\left\| {\bf{U}}_k \right\|_F^2  \le  0 , \; \forall k ,  \\
		\quad \Re \left\{ {{\text{Tr}}\left\{ {{{{\bf{\hat H}}}_{u,l,1}}{{{\bf{\tilde \Phi }}}_\imath}} \right\}} \right\} \ge \sqrt {\sigma _{\mathrm{c},u}^2{\Gamma _{u,l}}} \sin \Omega , \\
		\quad \Re \left\{ {{\text{Tr}}\left\{ {{{{\bf{\hat H}}}_{u,l,2}}{{{\bf{\tilde \Phi }}}_\imath}} \right\}} \right\} \ge \sqrt {\sigma _{\mathrm{c},u}^2{\Gamma _{u,l}}} \sin \Omega ,
	\end{numcases}
	\label{Neq:3_35}%
\end{subequations}
where ${{{\bf{\hat H}}}_{u,l,1}} = {{{\bf{\tilde H}}}_{u,l}}\left( {\sin \Omega  + {e^{ - \jmath \frac{\pi }{2}}}\cos \Omega } \right)$ and ${{{\bf{\hat H}}}_{u,l,2}} = {{{\bf{\tilde H}}}_{u,l}}\left( {\sin \Omega  - {e^{ - \jmath \frac{\pi }{2}}}\cos \Omega } \right)$.
Similar to problem $\mathcal{P}^2_{{\rm AL},{\bf W}}$, problem $\mathcal{P}^{2-2}_{{\text{AL}}, {\left\{ {\bm \Phi}_\imath \right\}}}$ is a convex SOCP and can be solved by IPM.

The proof is completed.

\section{Proof of {Theorem \ref{the:1}} }\label{Proof_the_1}
We start by rewriting objective \eqref{eq:3_40a} as \cite{boyd2004convex}
\begin{equation}
	\begin{aligned}
		& \mathbb{I}_g({\bm \Phi}_{g} , {\bm \Theta}_{g} , {\bm \Lambda}_g ) \\
		& =\Re  \left\{ {{\text{Tr}}\left( {{\bf{\Lambda }}_g^H\left( {{{\bf{\Phi }}_{g}} - {{\bf{\Theta }}_{g}}} \right)} \right)} \right\} + \frac{\varrho }{2}\left\| {{{\bf{\Phi }}_{g}} - {{\bf{\Theta }}_{g}}} \right\|_F^2 \\
        & = - \Re \left\{ {\rm Tr} \left( {{\bf{\Theta }}_{g}^H}{{{\left( {{{\bf{\Lambda }}_g} + {\varrho }{{\bf{\Phi }}_{g}}} \right)}} } \right) \right\} + \underbrace{\frac{\varrho }{2}\left\| {{{\bf{\Phi }}_{g}}} \right\|_F^2 + {\varrho }M}_{\text{constant}}.\notag
	\end{aligned}
	\label{eq:B_1}%
\end{equation}
Then, problem ${\mathcal{P}^{2-1}_{{\text{AL}}, {{\bm \Theta}_{g}}}}$ can be simplified as
\begin{equation}
	\begin{aligned}
		\mathop {\max }\limits_{{{\bf{\Phi }}_{g}}} &\;\;  \Re \left\{ {\rm Tr} \left( {{\bf{\Theta }}_{g}^H}{{{\left( {{{\bf{\Lambda }}_g} + {\varrho }{{\bf{\Phi }}_{g}}} \right)}} } \right) \right\}  \\
		{\rm {s.t.}} & \;\; {\bm \Theta}_{g}^H{\bm \Theta}_{g} = {\bf I}_{M}. 
	\end{aligned}
	\label{eq:B_2}
\end{equation}
Performing SVD to ${{{\bf{\Lambda }}_g} + {\varrho }{{\bf{\Phi }}_{g}}}$ as ${{\bf{B}}_g}{{\bf{\Sigma }}_g}{\bf{D}}_g^H = {{\bf{\Lambda }}_g} + {\varrho }{{\bf{\Phi }}_{g}}$, we can re-arrange the objective of \eqref{eq:B_2} as
\begin{equation}
	\begin{aligned}
		\Re & \left\{ {{\text{Tr}}\left( {{\bf{\Theta }}_{g}^H\left( {{{\bf{\Lambda }}_g} + {\varrho }{{\bf{\Phi }}_{g}}} \right)} \right)} \right\} \\
        & = \Re \left\{ {{\text{Tr}}\left( {{{\bf{\Sigma }}_g}{{\bf{Z}}_g}} \right)} \right\} =  \sum\limits_{i = 1}^M {{\bf{\Sigma }}_g}\left[ {i,i} \right]{{\bf{Z}}_g}\left[ {i,i} \right], 
	\end{aligned}
	\label{eq:B_3}
\end{equation}
where ${{\bf{Z}}_g} = {\bf{D}}_g^H{\bf{\Theta }}_{g}^H{{\bf{B}}_g}$.
\eqref{eq:B_3} achieves its maximum when ${{\bf{Z}}_g} = {\bf I}_{M \times 2M}$, yielding the optimal solution ${\bm \Theta}_{g} = {\bf{B}}_g\left[ {{{\bf{I}}_{M \times M}},{{\bf{0}}_{M \times M}}} \right]{{\bf{D}}_g^H}$. 

The proof is thus completed.

\section{Proof of {Proposition \ref{pro:1}} }\label{Proof_2}

The following proof employs a similar argument to that in \cite[Appendices]{Fan2018TSP}. 
Specifically, the first and second parts of this proof demonstrate the sufficient decrease property and the bounded property of the proposed algorithm, respectively. 
Building on these two critical properties, we establish that the optimal solution is achieved in the third part.

In the first part, we demonstrate the sufficient decreasing property of the proposed algorithm.
Specifically,
\begin{subequations}
	\begin{align}
		& {\mathcal{L}}\left( \left\{ \mathbf{U}_k^{n+1} \right\} , \mathbf{W}^n, \left\{ \mathbf{\Phi }_{g}^n \right\},   \left\{ {\bm \Theta}_{g}^n \right\} \right) \notag \\
		& \qquad \qquad - {\mathcal{L}}\left( \left\{ \mathbf{U}_k^n \right\} , \mathbf{W}^n, \left\{ \mathbf{\Phi }_{g}^n \right\},   \left\{ {\bm \Theta}_{g}^n \right\} \right) \\
		& \mathop  \le \limits^{(a)}  \min_k - \text{SCNR}_k(\mathbf{W}^n , \left\{ \mathbf{\Phi }_{g}^n \right\} , \mathbf{U}_k^{n+1}) \notag \\
		& \qquad \qquad \qquad + \text{SCNR}_k(\mathbf{W}^n , \left\{ \mathbf{\Phi }_{g}^n \right\} , \mathbf{U}_k^{n}) \mathop  \le \limits^{(b)} 0 ,
	\end{align}
	\label{eq:C1}%
\end{subequations}
where $(a)$ holds because updating $\{\mathbf{U}_k\}$ is independent of other variables; 
(b) holds because updating $\mathbf{U}_k$ achieves its optimum by \eqref{eq:3_22}, resulting in a non-increasing objective function.
For updating $\mathbf{W}$, we have
\begin{subequations}
	\begin{align}
		& {\mathcal{L}}\left( \left\{ \mathbf{U}_k^{n+1} \right\} , \mathbf{W}^{n+1}, \left\{ \mathbf{\Phi }_{g}^n \right\},   \left\{ {\bm \Theta}_{g}^n \right\} \right) \notag \\
		& \qquad \qquad - {\mathcal{L}}\left( \left\{ \mathbf{U}_k^{n+1} \right\} , \mathbf{W}^n, \left\{ \mathbf{\Phi }_{g}^n \right\},   \left\{ {\bm \Theta}_{g}^n \right\} \right) \\
		& =  \gamma_{n+1} - \gamma_n  \mathop  \le \limits^{(a)} 0 ,
	\end{align}
	\label{eq:C2}%
\end{subequations}
where $(a)$ holds because problem \eqref{eq:3_30} is a convex problem that can be optimally solved, thereby ensuring that updating $\mathbf{W}$ results in a non-increasing objective function.

Since updating $\left\{ \mathbf{\Phi }_{g} \right\}$ follows the same procedure as updating $\mathbf{W}$, we similarly have
\begin{equation}
	\begin{aligned}
		& {\mathcal{L}}\left( \left\{ \mathbf{U}_k^{n+1} \right\} , \mathbf{W}^{n+1}, \left\{ \mathbf{\Phi }_{g}^{n+1} \right\},   \left\{ {\bm \Theta}_{g}^n \right\} \right) \\
		& \qquad \qquad - {\mathcal{L}}\left( \left\{ \mathbf{U}_k^{n+1} \right\} , \mathbf{W}^{n+1}, \left\{ \mathbf{\Phi }_{g}^n \right\},   \left\{ {\bm \Theta}_{g}^n \right\} \right) \le  0 .
	\end{aligned}
	\label{eq:C3}
\end{equation}

Updating $\left\{ {\bm \Theta}_{g} \right\}$ using \textbf{\textit{Theorem \ref{the:1}}} yields the optimal solution, which guarantees
\begin{equation}
	\begin{aligned}
		& {\mathcal{L}}\left( \left\{ \mathbf{U}_k^{n+1} \right\} , \mathbf{W}^{n+1}, \left\{ \mathbf{\Phi }_{g}^{n+1} \right\},   \left\{ {\bm \Theta}_{g}^{n+1} \right\} \right) \\
		& \qquad \quad - {\mathcal{L}}\left( \left\{ \mathbf{U}_k^{n+1} \right\} , \mathbf{W}^{n+1}, \left\{ \mathbf{\Phi }_{g}^{n+1} \right\},   \left\{ {\bm \Theta}_{g}^n \right\} \right) \le  0 .
	\end{aligned}
	\label{eq:C4}
\end{equation}

By combining \eqref{eq:C1}-\eqref{eq:C4}, we conclude that the objective \eqref{eq:3_16c} is non-increasing after iterations, i.e.,
\begin{equation}
	\begin{aligned}
		& {\mathcal{L}}\left( \left\{ \mathbf{U}_k^{n+1} \right\} , \mathbf{W}^{n+1}, \left\{ \mathbf{\Phi }_{g}^{n+1} \right\},   \left\{ {\bm \Theta}_{g}^{n+1} \right\} \right) \\
		& \qquad \qquad - {\mathcal{L}}\left( \left\{ \mathbf{U}_k^{n} \right\} , \mathbf{W}^n, \left\{ \mathbf{\Phi }_{g}^n \right\},   \left\{ {\bm \Theta}_{g}^n \right\} \right) \le  0 .
	\end{aligned}
	\label{eq:C5}
\end{equation}

In the second part, we show the bounded property of the sequence $\left\{ \{ \mathbf{U}_k^n \} , \mathbf{W}^n, \{ \mathbf{\Phi }_{g}^n \},   \{ {\bm \Theta}_{g}^n \} \right\}$.
Since $\left\{ \mathbf{U}_k^n \right\}$ is updated by tackling the Rayleigh quotient in \eqref{eq:3_21}, the $\left\{ \mathbf{U}_k^n \right\}$ is always bounded.
Recalling that $\mathbf{W}^n$ is constrained by the available transmit power, $\mathbf{W}^n$ is bounded, i.e., $\left\| \mathbf{W}^n \right\|_F^2 \le E$.
Since $\left\{ {\bm \Theta}_{g}^n \right\}$ satisfies the orthogonal constraint, $\left\{ {\bm \Theta}_{g}^n \right\}$ must also satisfy $\left\| {\bm \Theta}_{g}^n \right\|_F^2 = M$, thereby ensuring boundedness.
Since $\lim\limits_{n\to\infty} {\bm \Lambda}_g^{n+1} - {\bm \Lambda}_g^{n} = \mathbf{0}$, along with the dual ascent step in \eqref{eq:3_19e}, we have
\begin{equation}
	\lim\limits_{n\to\infty} \mathbf{\Phi }_{g}^{n} - {\bm \Theta}_{g}^{n} = \mathbf{0},  g=1,\cdots,G.
	\label{neq:53}
\end{equation}
Thus, we have
\begin{equation}
	\left\|\mathbf{\Phi }_{g}^n \right\|_F^2 \le \left\| \mathbf{\Phi }_{g}^n - {\bm \Theta}_{g}^n \right\|_F^2 + \left\| {\bm \Theta}_{g}^n \right\|_F^2 , 
\end{equation}
which implies $\left\{ {\bm \Theta}_{g}^n \right\}$ is bounded, $\mathbf{\Phi }_{g}^n$ is also bounded.
Based on the above illustrations, we achieve the bounded property of the sequence $\left\{ \{ \mathbf{U}_k^n \} , \mathbf{W}^n, \{ \mathbf{\Phi }_{g}^n \},   \{ {\bm \Theta}_{g}^n \} \right\}$.

Furthermore, since every term of the sequence $\left\{ \left\{ \mathbf{U}_k^n \right\} , \mathbf{W}^n, \left\{ \mathbf{\Phi }_{g}^n \right\},   \left\{ {\bm \Theta}_{g}^n \right\} \right\}$ is bounded, the augmented Lagrangian ${\mathcal{L}}\left( \left\{ \mathbf{U}_k^n \right\} , \mathbf{W}^n, \left\{ \mathbf{\Phi }_{g}^n \right\},   \left\{ {\bm \Theta}_{g}^n \right\} \right)$ is also bounded.

In the third part, we show limit point $\left\{ \{ \bar{\mathbf{U}}_k \} , \bar{\mathbf{W}}, \{ \bar{\mathbf{\Phi }}_g \},   \{ \bar{{\bm \Theta}}_g \} \right\}$ of $\left\{ \{ \mathbf{U}_k^n \} , \mathbf{W}^n, \{ \mathbf{\Phi }_{g}^n \},   \{ {\bm \Theta}_{g}^n \} \right\}$ is a stationary point.
Specifically, based on the bounded property, there exists a limit point $\left\{ \{ \bar{\mathbf{U}}_k \} , \bar{\mathbf{W}}, \{ \bar{\mathbf{\Phi }}_g \},   \{ \bar{{\bm \Theta}}_g \} \right\}$ such that
\begin{equation}\label{neq:55}
	\begin{aligned}
		\lim\limits_{n\to\infty} {\mathbf{U}}_k^n = \bar{\mathbf{U}}_k, 
		&\lim\limits_{n\to\infty} {\mathbf{W}}^n = \bar{\mathbf{W}}, \\
		\lim\limits_{n\to\infty} {\mathbf{\Phi }}_g^n = \bar{\mathbf{\Phi }}_g,
		&\lim\limits_{n\to\infty} {{\bm \Theta}}_g^n = \bar{{\bm \Theta}}_g.
	\end{aligned}
\end{equation}
Hence, based on \eqref{neq:53} and \eqref{neq:55}, we have
\begin{equation}
	\lim\limits_{n\to\infty} \mathbf{\Phi }_{g}^{n} - {\bm \Theta}_{g}^{n} = \bar{\mathbf{\Phi }}_{g} - \bar{\bm \Theta}_{g} = \mathbf{0},  g=1,\cdots,G.
\end{equation}

Moreover, based on the sufficient decrease property and the bounded property of  ${\mathcal{L}}\left( \left\{ \mathbf{U}_k^n \right\} , \mathbf{W}^n, \left\{ \mathbf{\Phi }_{g}^n \right\},   \left\{ {\bm \Theta}_{g}^n \right\} \right)$ discussed in the first and second parts, we then have
\begin{equation}
	\begin{aligned}
		& \lim\limits_{n\to\infty} {\mathcal{L}}\left( \left\{ \mathbf{U}_k^n \right\} , \mathbf{W}^n, \left\{ \mathbf{\Phi }_{g}^n \right\},   \left\{ {\bm \Theta}_{g}^n \right\} \right) \\
		& \qquad \qquad = {\mathcal{L}}\left( \left\{ \bar{\mathbf{U}}_k \right\} , \bar{\mathbf{W}}, \left\{ \bar{\mathbf{\Phi }}_g \right\},   \left\{ \bar{{\bm \Theta}}_g \right\} \right) .
	\end{aligned}
\end{equation}
This implies that the limit point $\left\{ \{ \bar{\mathbf{U}}_k \} , \bar{\mathbf{W}}, \{ \bar{\mathbf{\Phi }}_g \},   \{ \bar{{\bm \Theta}}_g \} \right\}$ is a stationary point, which leads to a sub-optimal solution.

The proof is completed.

\footnotesize
\balance
\bibliographystyle{IEEEtran}
\bibliography{IEEEabrv,./ref_arv.bib}

% Generated by IEEEtran.bst, version: 1.14 (2015/08/26)
\begin{thebibliography}{10}
\providecommand{\url}[1]{#1}
\csname url@samestyle\endcsname
\providecommand{\newblock}{\relax}
\providecommand{\bibinfo}[2]{#2}
\providecommand{\BIBentrySTDinterwordspacing}{\spaceskip=0pt\relax}
\providecommand{\BIBentryALTinterwordstretchfactor}{4}
\providecommand{\BIBentryALTinterwordspacing}{\spaceskip=\fontdimen2\font plus
\BIBentryALTinterwordstretchfactor\fontdimen3\font minus
  \fontdimen4\font\relax}
\providecommand{\BIBforeignlanguage}[2]{{%
\expandafter\ifx\csname l@#1\endcsname\relax
\typeout{** WARNING: IEEEtran.bst: No hyphenation pattern has been}%
\typeout{** loaded for the language `#1'. Using the pattern for}%
\typeout{** the default language instead.}%
\else
\language=\csname l@#1\endcsname
\fi
#2}}
\providecommand{\BIBdecl}{\relax}
\BIBdecl

\bibitem{cui2021integrating}
Y.~Cui, F.~Liu, X.~Jing, and J.~Mu, ``Integrating sensing and communications
  for ubiquitous iot: Applications, trends, and challenges,'' \emph{{IEEE}
  Netw.}, vol.~35, no.~5, pp. 158--167, 2021.

\bibitem{nowak2016co}
M.~Nowak, M.~Wicks, Z.~Zhang, and Z.~Wu, ``Co-designed radar-communication
  using linear frequency modulation waveform,'' \emph{{IEEE} Aerosp. Electron.
  Syst. Mag.}, vol.~31, no.~10, pp. 28--35, 2016.

\bibitem{hassanien2016signaling}
A.~Hassanien, M.~G. Amin, Y.~D. Zhang, and F.~Ahmad, ``Signaling strategies for
  dual-function radar communications: An overview,'' \emph{{IEEE} Aerosp.
  Electron. Syst. Mag.}, vol.~31, no.~10, pp. 36--45, 2016.

\bibitem{wu2021frequency}
K.~Wu, J.~A. Zhang, X.~Huang, and Y.~J. Guo, ``Frequency-hopping {MIMO}
  radar-based communications: An overview,'' \emph{{IEEE} Aerosp. Electron.
  Syst. Mag.}, 2021.

\bibitem{kumari2017ieee}
P.~Kumari, J.~Choi, N.~Gonz{\'a}lez-Prelcic, and R.~W. Heath, ``{IEEE} 802.11
  ad-based radar: An approach to joint vehicular communication-radar system,''
  \emph{{IEEE} Trans. Veh. Technol.}, vol.~67, no.~4, pp. 3012--3027, 2017.

\bibitem{dokhanchi2019mmwave}
S.~H. Dokhanchi, B.~S. Mysore, K.~V. Mishra, and B.~Ottersten, ``A {mmWave}
  automotive joint radar-communications system,'' \emph{{IEEE} Trans. Aerosp.
  Electron. Syst.}, vol.~55, no.~3, pp. 1241--1260, 2019.

\bibitem{sturm2011waveform}
C.~Sturm and W.~Wiesbeck, ``Waveform design and signal processing aspects for
  fusion of wireless communications and radar sensing,'' \emph{Proc. {IEEE}},
  vol.~99, no.~7, pp. 1236--1259, 2011.

\bibitem{zhang2021overview}
J.~A. Zhang, F.~Liu, C.~Masouros, R.~W. Heath, Z.~Feng, L.~Zheng, and
  A.~Petropulu, ``An overview of signal processing techniques for joint
  communication and radar sensing,'' \emph{{IEEE} J. Sel. Topics Signal
  Process.}, vol.~15, no.~6, pp. 1295--1315, 2021.

\bibitem{liu2022integrated}
R.~Liu, M.~Li, H.~Luo, Q.~Liu, and A.~L. Swindlehurst, ``Integrated sensing and
  communication with reconfigurable intelligent surfaces: Opportunities,
  applications, and future directions,'' \emph{{IEEE} Wireless Commun.},
  vol.~30, no.~1, pp. 50--57, 2023.

\bibitem{cheng2023twc}
Z.~Cheng, L.~Wu, B.~Wang, M.~R.~B. Shankar, and B.~Ottersten,
  ``Double-phase-shifter based hybrid beamforming for {mmWave} {DFRC} in the
  presence of extended target and clutters,'' \emph{{IEEE} Trans. Wireless
  Commun.}, vol.~22, no.~6, pp. 3671--3686, 2023.

\bibitem{DiRenzo2020}
M.~Di~Renzo, A.~Zappone, M.~Debbah, M.-S. Alouini, C.~Yuen \emph{et~al.},
  ``Smart radio environments empowered by reconfigurable intelligent surfaces:
  How it works, state of research, and the road ahead,'' \emph{{IEEE} J. Sel.
  Areas Commun.}, vol.~38, no.~11, pp. 2450--2525, 2020.

\bibitem{SGong2019}
S.~Gong, X.~Lu, D.~T. Hoang, D.~Niyato, L.~Shu, D.~I. Kim, and Y.-C. Liang,
  ``Toward smart wireless communications via intelligent reflecting surfaces: A
  contemporary survey,'' \emph{{IEEE} Commun. Surveys Tuts.}, vol.~22, no.~4,
  pp. 2283--2314, 2020.

\bibitem{K-KWong}
K.-K. Wong, K.-F. Tong, Z.~Chu, and Y.~Zhang, ``A vision to smart radio
  environment: Surface wave communication superhighways,'' \emph{{IEEE}
  Wireless Commun.}, vol.~28, no.~1, pp. 112--119, 2020.

\bibitem{QWu2019}
Q.~Wu and R.~Zhang, ``Towards smart and reconfigurable environment: Intelligent
  reflecting surface aided wireless network,'' \emph{{IEEE} Commun. Mag.},
  vol.~58, no.~1, pp. 106--112, 2019.

\bibitem{liu2022joint}
R.~Liu, M.~Li, Y.~Liu, Q.~Wu, and Q.~Liu, ``Joint transmit waveform and passive
  beamforming design for {RIS}-aided {DFRC} systems,'' \emph{{IEEE} J. Sel.
  Topics Signal Process.}, vol.~16, no.~5, pp. 995--1010, 2022.

\bibitem{wei2022multi}
T.~Wei, L.~Wu, K.~V. Mishra, and M.~R. Bhavani~Shankar, ``{Multi-IRS}-aided
  doppler-tolerant wideband {DFRC} system,'' \emph{{IEEE} Trans. Commun.},
  vol.~71, no.~11, pp. 6561--6577, 2023.

\bibitem{yan2022reconfigurable}
S.~Yan, S.~Cai, W.~Xia, J.~Zhang, and S.~Xia, ``A reconfigurable intelligent
  surface aided dual-function radar and communication system,'' in \emph{Proc.
  IEEE Int. Symp. Joint Commun. Sens.}, vol. Mar., 2022, pp. 1--6.

\bibitem{sankar2022beamforming}
R.~P. Sankar, S.~P. Chepuri, and Y.~C. Eldar, ``Beamforming in integrated
  sensing and communication systems with reconfigurable intelligent surfaces,''
  \emph{{IEEE} Trans. Wireless Commun.}, vol.~23, no.~5, pp. 4017--4031, 2024.

\bibitem{liu2023snr}
R.~Liu, M.~Li, Q.~Liu, and A.~L. Swindlehurst, ``{SNR/CRB}-constrained joint
  beamforming and reflection designs for {RIS-ISAC} systems,'' \emph{{IEEE}
  Trans. Wireless Commun.}, vol.~23, no.~7, pp. 7456--7470, 2024.

\bibitem{song2022cram}
X.~Song, T.~X. Han, and J.~Xu, ``Cram\'er-rao bound minimization for
  {IRS}-enabled multiuser integrated sensing and communication with extended
  target,'' \emph{{IEEE} Trans. Wireless Commun.}, pp. 1--1, 2024.

\bibitem{Hua2022Joint}
M.~Hua, Q.~Wu, C.~He, S.~Ma, and W.~Chen, ``Joint active and passive
  beamforming design for {IRS}-aided radar-communication,'' \emph{{IEEE} Trans.
  Wireless Commun.}, vol.~22, no.~4, pp. 2278--2294, 2023.

\bibitem{wang2021joint}
X.~Wang, Z.~Fei, J.~Huang, and H.~Yu, ``Joint waveform and discrete phase shift
  design for {RIS}-assisted integrated sensing and communication system under
  cram\'er-rao bound constraint,'' \emph{{IEEE} Trans. Veh. Technol.}, vol.~71,
  no.~1, pp. 1004--1009, 2021.

\bibitem{sankar2021joint}
R.~P. Sankar, B.~Deepak, and S.~P. Chepuri, ``Joint communication and radar
  sensing with reconfigurable intelligent surfaces,'' in \emph{Proc. IEEE 22nd
  Int. Workshop Signal Process. Adv. Wireless Commun. (SPAWC)}, 2021, pp.
  471--475.

\bibitem{JXuHybridRIS}
J.~Xu, Y.~Liu, X.~Mu, and O.~A. Dobre, ``{STAR-RISs}: Simultaneous transmitting
  and reflecting reconfigurable intelligent surfaces,'' \emph{{IEEE} Commun.
  Lett.}, vol.~25, no.~9, pp. 3134--3138, 2021.

\bibitem{HZhang}
H.~Zhang, S.~Zeng, B.~Di, Y.~Tan, M.~Di~Renzo, M.~Debbah, Z.~Han, H.~V. Poor,
  and L.~Song, ``Intelligent omni-surfaces for full-dimensional wireless
  communications: Principles, technology, and implementation,'' \emph{{IEEE}
  Commun. Mag.}, vol.~60, no.~2, pp. 39--45, 2022.

\bibitem{wang2022stars}
Z.~Wang, X.~Mu, and Y.~Liu, ``{STARS} enabled integrated sensing and
  communications,'' \emph{{IEEE} Trans. Wireless Commun.}, vol.~22, no.~10, pp.
  6750--6765, 2023.

\bibitem{meng2022sensing}
K.~Meng, Q.~Wu, W.~Chen, and D.~Li, ``Sensing-assisted communication in
  vehicular networks with intelligent surface,'' \emph{{IEEE} Trans. Veh.
  Technol.}, vol.~73, no.~1, pp. 876--893, 2024.

\bibitem{zhang2023star}
Z.~Zhang, Y.~Liu, Z.~Wang, and J.~Chen, ``{STARS-ISAC}: How many sensors do we
  need?'' \emph{{IEEE} Trans. Wireless Commun.}, vol.~23, no.~2, pp.
  1085--1099, 2024.

\bibitem{SShen}
S.~Shen, B.~Clerckx, and R.~Murch, ``Modeling and architecture design of
  reconfigurable intelligent surfaces using scattering parameter network
  analysis,'' \emph{{IEEE} Trans. Wireless Commun.}, vol.~21, no.~2, pp.
  1229--1243, 2021.

\bibitem{li2022beyondtwc}
H.~Li, S.~Shen, and B.~Clerckx, ``Beyond diagonal reconfigurable intelligent
  surfaces: From transmitting and reflecting modes to single-, group-, and
  fully-connected architectures,'' \emph{{IEEE} Trans. Wireless Commun.},
  vol.~22, no.~4, pp. 2311--2324, 2023.

\bibitem{nerini2022optimal}
M.~Nerini, S.~Shen, and B.~Clerckx, ``Closed-form global optimization of beyond
  diagonal reconfigurable intelligent surfaces,'' \emph{{IEEE} Trans. Wireless
  Commun.}, vol.~23, no.~2, pp. 1037--1051, 2024.

\bibitem{li2022beyond_MSM}
H.~Li, S.~Shen, and B.~Clerckx, ``Beyond diagonal reconfigurable intelligent
  surfaces: A multi-sector mode enabling highly directional full-space wireless
  coverage,'' \emph{{IEEE} J. Sel. Areas Commun.}, vol.~41, no.~8, pp.
  2446--2460, 2023.

\bibitem{LQC2022NDRIS}
Q.~Li, M.~El-Hajjar, I.~Hemadeh, A.~Shojaeifard, A.~A.~M. Mourad, B.~Clerckx,
  and L.~Hanzo, ``Reconfigurable intelligent surfaces relying on non-diagonal
  phase shift matrices,'' \emph{{IEEE} Trans. Veh. Technol.}, vol.~71, no.~6,
  pp. 6367--6383, 2022.

\bibitem{Li2020Tutorial}
A.~Li, D.~Spano, J.~Krivochiza, S.~Domouchtsidis, C.~G. Tsinos, C.~Masouros,
  S.~Chatzinotas, Y.~Li, B.~Vucetic, and B.~Ottersten, ``A tutorial on
  interference exploitation via symbol-level precoding: Overview,
  state-of-the-art and future directions,'' \emph{{IEEE} Commun. Surveys
  Tuts.}, vol.~22, no.~2, pp. 796--839, 2020.

\bibitem{Li2018Interference}
A.~Li and C.~Masouros, ``Interference exploitation precoding made practical:
  Optimal closed-form solutions for {PSK} modulations,'' \emph{{IEEE} Trans.
  Wireless Commun.}, vol.~17, no.~11, pp. 7661--7676, 2018.

\bibitem{shao2022target}
X.~Shao, C.~You, W.~Ma, X.~Chen, and R.~Zhang, ``Target sensing with
  intelligent reflecting surface: Architecture and performance,'' \emph{{IEEE}
  J. Sel. Areas Commun.}, vol.~40, no.~7, pp. 2070--2084, 2022.

\bibitem{li2008mimo}
J.~Li and P.~Stoica, \emph{{MIMO} radar signal processing}.\hskip 1em plus
  0.5em minus 0.4em\relax John Wiley \& Sons, 2008.

\bibitem{cheng2018spectrally}
Z.~Cheng, B.~Liao, Z.~He, Y.~Li, and J.~Li, ``Spectrally compatible waveform
  design for {MIMO} radar in the presence of multiple targets,'' \emph{{IEEE}
  Trans. Signal Process.}, vol.~66, no.~13, pp. 3543--3555, 2018.

\bibitem{cui2013mimo}
G.~Cui, H.~Li, and M.~Rangaswamy, ``{MIMO} radar waveform design with constant
  modulus and similarity constraints,'' \emph{{IEEE} Trans. Signal Process.},
  vol.~62, no.~2, pp. 343--353, 2013.

\bibitem{de2008design}
A.~De~Maio, S.~De~Nicola, Y.~Huang, Z.-Q. Luo \emph{et~al.}, ``Design of phase
  codes for radar performance optimization with a similarity constraint,''
  \emph{{IEEE} Trans. Signal Process.}, vol.~57, no.~2, pp. 610--621, 2008.

\bibitem{boyd2004convex}
S.~Boyd, S.~P. Boyd, and L.~Vandenberghe, \emph{Convex optimization}.\hskip 1em
  plus 0.5em minus 0.4em\relax Cambridge university press, 2004.

\bibitem{aubry2021reconfigurable}
A.~Aubry, A.~De~Maio, and M.~Rosamilia, ``Reconfigurable intelligent surfaces
  for {N-LOS} radar surveillance,'' \emph{{IEEE} Trans. Veh. Technol.},
  vol.~70, no.~10, pp. 10\,735--10\,749, 2021.

\bibitem{buzzi2021radar}
S.~Buzzi, E.~Grossi, M.~Lops, and L.~Venturino, ``Radar target detection aided
  by reconfigurable intelligent surfaces,'' \emph{{IEEE} Signal Process.
  Lett.}, vol.~28, pp. 1315--1319, 2021.

\bibitem{lu2021target}
W.~Lu, Q.~Lin, N.~Song, Q.~Fang, X.~Hua, and B.~Deng, ``Target detection in
  intelligent reflecting surface aided distributed {MIMO} radar systems,''
  \emph{{IEEE} Sens. Lett.}, vol.~5, no.~3, pp. 1--4, 2021.

\bibitem{zhang2017matrix}
X.-D. Zhang, \emph{Matrix analysis and applications}.\hskip 1em plus 0.5em
  minus 0.4em\relax Cambridge University Press, 2017.

\bibitem{Fan2018TSP}
W.~Fan, J.~Liang, and J.~Li, ``Constant modulus {MIMO} radar waveform design
  with minimum peak sidelobe transmit beampattern,'' \emph{{IEEE} Trans. Signal
  Process.}, vol.~66, no.~16, pp. 4207--4222, 2018.

\end{thebibliography}

\end{document}